\documentclass[11pt]{article}
\usepackage[margin=1in]{geometry}
\usepackage{authblk}
\usepackage{bm}
\usepackage{amsmath}
\usepackage{graphicx}
\usepackage{xcolor}
\usepackage[T1]{fontenc}
\usepackage{newtxtext,newtxmath}
\usepackage{appendix}
\usepackage{placeins}
\usepackage{float}
\usepackage{cite} 
\graphicspath{{figures/}}
\usepackage{hyperref}
\hypersetup{colorlinks=true,linkcolor=blue,citecolor=blue,urlcolor=blue}

\usepackage{indentfirst}

\AtBeginDocument{%
    \setlength{\parindent}{2em}
    \setlength{\parskip}{0pt}
}

\begin{document}

\title{Quasi-equilibrium tidal folds of white dwarfs around rotating black holes in shift-symmetric Einstein--scalar--Gauss--Bonnet gravity}
\author[1]{Shoupan Liu\thanks{shoupan\_liu@sjtu.edu.cn}}
\author[1]{Meilin Liu\thanks{meilin.liu@sjtu.edu.cn (corresponding author)}}
\author[1,2]{Yu Wang\thanks{sjtu2686361@sjtu.edu.cn}}
\author[3]{Yunqi Liu\thanks{yunqiliu@yzu.edu.cn}}
\affil[1]{School of Aeronautics and Astronautics, Shanghai Jiao Tong University, Shanghai 200240, China}
\affil[2]{School of Physics and Astronomy, Shanghai Jiao Tong University, Shanghai 200240, China}
\affil[3]{Center for Gravitation and Cosmology, College of Physical Science and Technology, Yangzhou University, Yangzhou 225009, China}
\date{} 

    \maketitle
    \begin{abstract}
We examine how the geometry of rotating black holes in linear,
shift-symmetric Einstein--scalar--Gauss--Bonnet (EsGB) gravity changes
the limiting tidal equilibria of white dwarfs (WDs). The relativistic
tidal field is obtained from numerical EsGB backgrounds in a
parallel-transported frame and coupled to a compressible affine model.
Comparisons with Kerr black holes matched in Arnowitt--Deser--Misner
(ADM) mass and angular momentum show a positive shift in the orbital
frequency at the tidal fold. At fixed spin the relative correction
follows an approximately quadratic dependence on the EsGB coupling,
consistent with the leading response of the metric to the scalar field.
Over the sampled backgrounds, the relative correction reaches order
$10^{-3}$. Black hole rotation also changes the relation between the
tidal fold and the innermost stable circular orbit (ISCO): the sampled
prograde folds remain outside the ISCO, while retrograde folds move
inside it at higher spins. Orbital orientation therefore controls
whether tidal equilibrium is lost before the orbital stability limit
is reached. For the stellar and black hole masses considered here,
the characteristic gravitational wave (GW) frequencies fall in the
millihertz band relevant to planned space based interferometers.
Future GW observations capable of resolving deviations from Kerr
predictions could constrain the EsGB coupling.
    \end{abstract}

\section{Introduction}\label{Introduction}

Tidal interactions between compact stars and black holes provide a setting in which stellar structure is coupled directly to the curvature of a strong gravitational field \cite{Hills1975,Rees1988,StoneEtAl2020,Gezari2021}. WDs are of particular interest: for suitable black hole masses and spins, they may undergo substantial tidal deformation before crossing the event horizon or reaching the final plunge \cite{RosswogEtAl2009,Kesden2012,GaftonRosswog2019}. Encounters between WDs and black holes at intermediate masses and at the lower end of the massive black hole range have consequently been investigated as possible sources of gravitational radiation and electromagnetic transients \cite{MaguireEtAl2020,RossiEtAl2021,HaasEtAl2012}. The character of any subsequent disruption, however, depends on the stellar equation of state (EOS), orbital configuration, relativistic dynamics, and the evolution of the stripped material \cite{PhysRevD.90.064020,GuillochonRamirezRuiz2013,StoneSariLoeb2013}.

These systems also motivate a connection between tidal physics and observations. A WD spiralling towards a massive black hole may emit gravitational waves (GWs) at low frequencies before strong tidal deformation or mass transfer, while subsequent debris accretion can, in favorable circumstances, produce electromagnetic emission \cite{SesanaEtAl2008,MaguireEtAl2020,CasalvieriFerrariStavridis2006}. Transients at high energies, including scenarios involving gamma-ray bursts, have been discussed in the broader literature on the disruption of compact stars, but they require additional assumptions about disruption, accretion, and outflows \cite{Wiggins_2000,HaasEtAl2012}. Of more direct relevance to a quasi-circular tidal calculation is the frequency scale associated with the end of an approximately adiabatic inspiral. Studies of systems containing a black hole and a neutron star have related tidal limits to possible GW spectral cutoffs, although fully dynamical simulations show that the actual cutoff need not coincide with the frequency estimated from a quasi-equilibrium sequence \cite{PhysRevD.81.064026,ShibataTaniguchi2008}. A tidal equilibrium limit can therefore identify a characteristic frequency without, by itself, predicting a waveform termination.

The relativistic tidal field experienced by an orbiting star is determined by projections of the Riemann tensor onto a local frame carried by its center of mass. Its effect on the star depends both on the external spacetime geometry and on the competition between self-gravity, pressure, and internal fluid motion \cite{PressTeukolsky1977,CarterLuminet1982}. Relativistic Roche models and compressible affine approximations provide a tractable description of this interplay \cite{Fishbone1973,CarterLuminet1985,LRS1,10.1093mnras,1986ApJS...61..219L,Ferrari_2009,PhysRevD.85.044045}. In particular, Wiggins and Lai constructed quasi-equilibrium sequences of compressible, possibly nonsynchronized stars moving on circular orbits around Kerr black holes \cite{Wiggins_2000}; higher-order relativistic tidal effects have also been investigated \cite{IshiiShibataMino2005}. Maselli, Kokkotas, and Laguna subsequently applied an affine model to a parametrized non-Kerr geometry, demonstrating how deviations in the tidal field can affect WD equilibria \cite{PhysRevD.93.064075}. Together, these studies establish the affine framework and motivate a more specific question: how does the critical tidal equilibrium respond when the rotating black hole geometry is determined by the field equations of a modified gravity theory?

A suitable example is shift-symmetric EsGB gravity with a linear scalar coupling to the Gauss--Bonnet invariant \cite{Sotiriou2015}. In this theory, spacetime curvature sources a nontrivial scalar field around a black hole, and the corresponding metric can depart from the Kerr solution even when its asymptotic mass and angular momentum are fixed \cite{PhysRevLett.112.251102,PhysRevD.90.124063,HuiNicolis2013,KleihausEtAl2011}. Stationary rotating black holes in the linearly coupled theory have been constructed numerically, and spectral approaches provide a systematic way of obtaining and testing such backgrounds \cite{DelgadoEtAl2020,Fernandes_2023,SullivanEtAl2021,FernandesMulryneDelgado2022}. This linear, shift-symmetric coupling should be distinguished from nonlinear EsGB models that exhibit spontaneous scalarization \cite{SilvaEtAl2018}. These rotating solutions make it possible to trace changes in the local tidal curvature to the underlying scalar coupling and then to the limiting equilibrium of an orbiting WD.

The central question is how the curvature produced by the scalar
coupling changes the limiting equilibrium of a WD on a rotating black hole
background. The WD responds to the Riemann tensor projected along its
orbit, so the scalar modification enters both the tidal forces and
the orbital and frame rotation rates that control the affine balance.
A second question is how this tidal limit competes with the ISCO,
which governs the radial stability of circular geodesics. These are
distinct physical thresholds and their ordering can depend on the
sense of orbital motion.

For a fixed stellar model, the relevant black hole parameters are the ADM
mass $M_{\rm ADM}$, the dimensionless spin
$\chi=J/M_{\rm ADM}^{2}$, and the dimensionless EsGB coupling
$\zeta=\alpha/M_{\rm ADM}^{2}$. To isolate the effect of the scalar
coupling, we compare every EsGB background with a Kerr solution
matched in its ADM mass and angular momentum \cite{Kerr1963}.
The comparison is carried out separately on prograde and retrograde
circular orbit branches.

We compute the tidal tensor in a parallel-transported frame along
equatorial circular geodesics and use it in a compressible affine
model for a cold, irrotational $n=3/2$ polytropic WD
\cite{LRS1,Wiggins_2000}. The limiting configurations are identified
as ordinary folds of the static equilibrium equations, and their
frequencies are compared with Kerr values at the same mass, spin, and
orbital orientation. We also compare the tidal fold radius with the
ISCO radius on each numerical background.

The computed EsGB--Kerr differences reveal three related physical
features. First, the relative fold frequency correction is positive
and approximately quadratic in the coupling at weak coupling, as
expected when the scalar field is sourced at first order but changes
the geometry at second order. Second, the magnitude of this correction
varies with spin, demonstrating that the response of the stellar tidal
limit depends on both the scalar coupling and rotation. Third, the
fold and ISCO have different spin dependences: the tidal fold appears
first along the sampled stable prograde sequences, whereas at higher
retrograde spins the ISCO precedes the static fold. This identifies
orbital orientation as a control on which instability limit is met
first.

The corresponding fold frequencies also provide an observationally
relevant scale. For a $0.40M_\odot$ WD around a
$2\times10^5M_\odot$ black hole, the dominant quadrupolar GW harmonic lies
near $18\,\mathrm{mHz}$ in the source frame, within the frequency
range of the Laser Interferometer Space Antenna (LISA)
\cite{LISAColpi2024} and of the TianQin mission
\cite{LuoTianQin2016}. If future GW measurements resolve the corresponding difference
from Kerr predictions after accounting for source parameters, they
could constrain the EsGB coupling and test gravity in the strong
field regime.

The paper is organized as follows. Section~\ref{Theoretical setup} specifies the EsGB theory and the construction of rotating black hole backgrounds. Section~\ref{Sec.3} develops the tidal field along circular orbits, affine stellar equations, and criterion for a static fold. Section~\ref{sec:numerical_results} presents the matched EsGB--Kerr sequences and their coupling and spin dependence. Technical derivations and normalization conventions are collected in the appendices. We use the metric signature $(-,+,+,+)$ and geometrized units with $G=c=1$, restoring physical constants where necessary.

    \section{Rotating EsGB black hole backgrounds}
\label{Theoretical setup}

\subsection{Field equations}
The linearly coupled, shift-symmetric EsGB model considered here is defined by the action
\begin{equation}\label{Act}
 S=\frac{1}{16\pi}\int d^4x\sqrt{-g}
 \left[R-\nabla_\mu\phi\nabla^\mu\phi+\alpha\phi\mathcal G\right],
\end{equation}
where the scalar field $\phi$ is dimensionless, whereas the coupling $\alpha$ has dimensions of squared length. The Gauss--Bonnet scalar appearing in the action is
\begin{equation}\label{GB}
 \mathcal G=R^2-4R_{\mu\nu}R^{\mu\nu}
 +R_{\mu\nu\rho\sigma}R^{\mu\nu\rho\sigma}.
\end{equation}
Variation with respect to the metric and the scalar field gives \cite{PhysRevD.103.044040,RipleyPretorius2019}
\begin{align}
 G_{\mu\nu}&=\nabla_\mu\phi\nabla_\nu\phi
 -\frac12 g_{\mu\nu}(\nabla\phi)^2
 -4\alpha P_{\mu\rho\nu\sigma}\nabla^\rho\nabla^\sigma\phi,
 \label{Modified Einstein Eq}\\
 \Box\phi&=-\frac12\alpha\mathcal G,
 \label{scalar field Eq}
\end{align}
where the curvature tensor $P_{\mu\rho\nu\sigma}$ is defined by
\begin{equation}\label{double dual Riemann tensor}
 P_{\mu\rho\nu\sigma}=R_{\mu\rho\nu\sigma}
 -2g_{\mu[\nu}R_{\sigma]\rho}
 +2g_{\rho[\nu}R_{\sigma]\mu}
 +Rg_{\mu[\nu}g_{\sigma]\rho}.
\end{equation}

The scalar coupling changes the spacetime metric and hence the Riemann curvature that determines the tidal field. To quantify its effect on the WD without prescribing a phenomenological deformation of Kerr, we construct stationary, axisymmetric EsGB black hole solutions directly from these field equations. Their metric functions supply the orbital and curvature information needed for the tidal calculation, while matching the ADM mass and spin provides a Kerr reference for the comparison.

\subsection{Numerical black hole construction}
Following Ref.~\cite{Fernandes_2023}, we adopt the quasi-isotropic ansatz for rotating black holes with equatorial reflection symmetry,
\begin{equation}
 ds^2=-f\mathcal N^2dt^2+\frac{gh}{f}(dr^2+r^2d\theta^2)
 +\frac{gr^2\sin^2\theta}{f}
 \left[d\varphi-\frac{(1-\mathcal N)W}{r}dt\right]^2.
 \label{LineElement}
\end{equation}
Here $\mathcal N=1-r_H/r$, where $r_H$ is the horizon coordinate radius. The metric functions $f,g,h,W$ and the scalar field $\phi$ depend on $(r,\theta)$. Equatorial reflection symmetry allows us to restrict the computational domain to $0\leq\theta\leq\pi/2$, with the radial coordinate extending from $r=r_H$ to infinity.

Asymptotic flatness requires $f,g,h\to1$ and $W,\phi\to0$ as $r\to\infty$. More specifically, the leading asymptotic terms are
\begin{align}
 f&=1-\frac{2(M_{\rm ADM}-r_H)}{r}+O(r^{-2}),\notag\\
 W&=\frac{2J}{r_Hr}+O(r^{-2}),
 &\phi&=\frac{Q_s}{r}+O(r^{-2}).
 \label{eq:ADM_asymptotics}
\end{align}
These asymptotic coefficients determine the ADM mass $M_{\rm ADM}$, angular momentum $J$, and scalar charge $Q_s$. Their extraction formulas, together with the definition of the dimensionless spin $\chi$, are
\begin{align}
 M_{\rm ADM}&=r_H+\lim_{r\to\infty}\frac{r^2}{2}\partial_rf,
 &J&=-\lim_{r\to\infty}\frac{r_Hr^2}{2}\partial_rW,\notag\\
 Q_s&=\lim_{r\to\infty}r\phi,
 &\chi&=\frac{J}{M_{\rm ADM}^2}.
 \label{eq:ADM_definitions}
\end{align}

To treat the unbounded radial domain spectrally, we introduce the compact coordinate $X=1-2r_H/r$, which maps the horizon to $X=-1$ and spatial infinity to $X=1$ \cite{Boyd2001,Trefethen2000,GrandclementNovak2009}. We then expand each of the five fields $\mathcal F\in\{f,g,h,W,\phi\}$ in radial Chebyshev polynomials and even angular cosine modes,
\begin{equation}
 \mathcal F(X,\theta)=\sum_{p=0}^{N_r}\sum_{q=0}^{N_\theta}
 \mathcal F_{pq}T_p(X)\cos(2q\theta).
 \label{eq:spectral_expansion}
\end{equation}
The spectral representation is supplemented by boundary conditions at the endpoints of the compactified interval and on the rotation axis. At the horizon ($X=-1$), regular near-horizon expansions of the field equations yield
\begin{align}
 f-2f_{,X}&=0,\qquad g+2g_{,X}=0,\notag\\
 h_{,X}&=0,\quad W-W_{,X}=0,\quad\phi_{,X}=0.
 \label{eq:horizon_BCs}
\end{align}
At spatial infinity ($X=1$), Eq.~\eqref{eq:ADM_asymptotics} requires $f=g=h=1$ and $\phi=0$; prescribing the dimensionless spin additionally imposes $W_{,X}+(1+f_{,X})^2\chi=0$. On the rotation axis, elementary flatness requires $h(X,0)=1$. With these boundary conditions, we solve the spectrally discretized field equations by Newton iteration, following Refs.~\cite{Fernandes_2023,SullivanEtAl2021}.

\section{White dwarf tidal equilibria}
    \label{Sec.3}

\subsection{Circular orbits and the relativistic tidal field}
\label{Sec.3 A}
The relativistic tidal field experienced by the WD is determined by its
center of mass orbit in the black hole geometry constructed above. In the test
body approximation, justified by $m/M_{\rm ADM}\ll 1$, we take this
orbit to be a circular equatorial geodesic, with four-velocity
 $u^\mu=u^t(1,0,0,\Omega_{\rm orb})$ and
 $\Omega_{\rm orb}=d\varphi/dt$. The radial geodesic equation
 determines the prograde and retrograde orbital branches \cite{BardeenPressTeukolsky1972,Carter1968,Wilkins1972},
\begin{equation}
 \Omega_{\rm orb}^{\pm}
 =\frac{-\partial_r g_{t\varphi}\pm
   \sqrt{(\partial_r g_{t\varphi})^2
              -(\partial_r g_{tt})(\partial_r g_{\varphi\varphi})}}
        {\partial_r g_{\varphi\varphi}}.
 \label{eq:circular_branches}
\end{equation}
For the nonnegative spin backgrounds considered here, the ``$+$''
and ``$-$'' branches correspond to prograde and retrograde orbits,
respectively, when $\chi>0$. In the nonrotating limit the two senses
of orbital motion are equivalent. The timelike normalization
$u^\mu u_\mu=-1$ then fixes
\begin{equation}
 u^t=\left[-\left(g_{tt}+2g_{t\varphi}\Omega_{\rm orb}
                 +g_{\varphi\varphi}\Omega_{\rm orb}^2\right)\right]^{-1/2}.
 \label{eq:circular_ut}
\end{equation}
We introduce an orbit-adapted orthonormal tetrad
$\widetilde\lambda^\mu{}_{\hat a}$ with
$\widetilde\lambda^\mu{}_{\hat0}=u^\mu$. Its spatial legs are
\begin{equation}\label{eq:orbit_adapted_tetrad}
 \begin{gathered}
 \widetilde\lambda^\mu{}_{\hat1}=\frac{\delta^\mu_r}{\sqrt{g_{rr}}},\quad
 \widetilde\lambda^\mu{}_{\hat2}=\frac{\delta^\mu_\theta}{\sqrt{g_{\theta\theta}}},\quad
 \widetilde\lambda^\mu{}_{\hat3}=
 \frac{u_\varphi\delta^\mu_t-u_t\delta^\mu_\varphi}
 {\sqrt{g_{t\varphi}^2-g_{tt}g_{\varphi\varphi}}}.
 \end{gathered}
\end{equation}
Although this tetrad is comoving, its radial and azimuthal legs are
not parallel transported along the circular orbit
\cite{10.1098/rspa.1983.0021,Chicone_2006}. Their rotation can be
compensated by a spatial rotation through an angle $\xi(\tau)$ in the
radial and azimuthal plane. The parallel-transported tetrad
$\lambda^\mu{}_{\hat a}$ therefore has $\lambda_{\hat0}=u$,
$\lambda_{\hat2}=\widetilde\lambda_{\hat2}$, and
\begin{align}
 \lambda^\mu{}_{\hat1}
  &=\cos\xi\,\widetilde\lambda^\mu{}_{\hat1}
    -\sin\xi\,\widetilde\lambda^\mu{}_{\hat3},\notag\\
 \lambda^\mu{}_{\hat3}
  &=\sin\xi\,\widetilde\lambda^\mu{}_{\hat1}
    +\cos\xi\,\widetilde\lambda^\mu{}_{\hat3}.
 \label{eq:parallel_tetrad_rotation}
\end{align}
Imposing $\nabla_u\lambda_{\hat i}=0$ and using the condition for a circular geodesic
gives
\begin{equation}\label{eq:parallel_frame_rate}
 \frac{d\xi}{d\tau}=-\widetilde\lambda_{\hat1\mu}
 u^\nu\nabla_\nu\widetilde\lambda^\mu{}_{\hat3}
 =\frac{\partial_r g_{t\varphi}+\Omega_{\rm orb}\partial_r g_{\varphi\varphi}}
 {2\sqrt{g_{rr}(g_{t\varphi}^2-g_{tt}g_{\varphi\varphi})}}.
\end{equation}
Unlike $\Omega_{\rm orb}=d\varphi/dt$, the frame rotation rate
is defined with respect to the WD proper time. The electric tidal tensor is obtained by projecting the Riemann
curvature onto the parallel-transported frame
\cite{Misner:1973prb,PoissonPoundVega2011}:
\begin{equation}\label{tidal tensor}
 \mathcal E_{\hat i\hat j}=R_{\mu\nu\rho\sigma}
 \lambda^\mu{}_{\hat0}\lambda^\nu{}_{\hat i}
 \lambda^\rho{}_{\hat0}\lambda^\sigma{}_{\hat j}.
\end{equation}
The corresponding geodesic deviation equation is
$D^2 n^{\hat i}/d\tau^2=-\mathcal E^{\hat i}{}_{\hat j}n^{\hat j}$
\cite{Pirani:1956tn,Mashhoon1975}. For the circular equatorial orbits considered here, the electric tidal tensor is diagonal in the orbit-adapted frame owing to equatorial reflection symmetry. Its principal directions describe how nearby freely falling fluid elements are stretched or compressed by the external curvature. Determining the resulting stellar shape additionally requires the balance with pressure, self-gravity, and internal motion, which we introduce next.

    \subsection{Affine model and quasi-equilibrium configurations}

The tidal tensor in Eq.~\eqref{tidal tensor} specifies the external curvature force, while the deformation of the WD also depends on its restoring forces. We describe this response with the compressible affine model \cite{LRS1,Wiggins_2000,PhysRevD.93.064075}, in which the density surfaces are approximated by similar ellipsoids with principal semi-axes $a_1,a_2,a_3$ \cite{Chandrasekhar1969}. We take $a_2$ to be perpendicular to the orbital plane; in the aligned configuration considered below, $a_1$ and $a_3$ lie in the radial and tangential directions. Here $R_0$ denotes the radius of the isolated, spherical reference WD. The model retains Newtonian WD gravity, appropriate when $m/R_0\ll1$, and assumes minimal matter coupling. We also neglect the WD's induced scalar response and its effect on the external geometry and orbit.

To isolate the effect of black hole geometry in the EsGB--Kerr comparison, we keep the stellar model fixed. Specifically, we adopt a cold, nonrelativistic electron degeneracy polytrope with $n=3/2$, which provides a spherical reference configuration for the compressible affine approximation \cite{Chandrasekhar1931,ShapiroTeukolsky1983}. We set $\mu_e=2$, appropriate for fully ionized helium or carbon and oxygen matter. The corresponding equation of state (EOS) is
\begin{equation}
 P=K\rho^{1+1/n}=K\rho^{5/3},\qquad
 K=\frac{(3\pi^2)^{2/3}\hbar^2}{5m_e(\mu_em_u)^{5/3}}.
 \label{eq:WD_EOS}
\end{equation}
Here $\rho$ is the mass density, $\hbar$ is the reduced Planck constant, $m_e$ is the electron mass, $m_u$ is the atomic mass unit, and $\mu_em_u$ is the mean mass per electron. For a specified WD mass and composition, its reference radius $R_0$ and the structural constants required by the affine model follow from the Lane--Emden equation \cite{Chandrasekhar1939}. With central mass density $\rho_c$, we write $\rho=\rho_c\Theta^n$, where $\Theta$ is the Lane--Emden function and $\xi_{\rm LE}$ is its dimensionless radial coordinate. The structure equations are
\begin{align}
 \frac{1}{\xi_{\rm LE}^2}\frac{d}{d\xi_{\rm LE}}
 \left(\xi_{\rm LE}^2\frac{d\Theta}{d\xi_{\rm LE}}\right)&=-\Theta^n,\notag\\
 R_0&=a_{\rm LE}\xi_1,\notag\\
 m&=4\pi a_{\rm LE}^3\rho_c\nu_n,\qquad
 \nu_n=-\xi_1^2\Theta'(\xi_1).
 \label{eq:Lane_Emden}
\end{align}
Here $a_{\rm LE}^2=(n+1)K\rho_c^{1/n-1}/(4\pi G)$, $\Theta(0)=1$, $\Theta'(0)=0$, and $\xi_1$ is the first zero of $\Theta$. These EOS and Lane--Emden relations are evaluated in physical units. The resulting density profile also determines the affine structural constants,
\begin{equation}
 \kappa_n=\frac{5}{3\xi_1^2\nu_n}\int_0^{\xi_1}
 \Theta^n\xi_{\rm LE}^4\,d\xi_{\rm LE},
 \qquad q_n=\kappa_n(1-n/5).
 \label{eq:structure_constants}
\end{equation}
For the representative low mass WD adopted here, $m=0.40M_\odot$ and $\mu_e=2$ give $R_0\simeq1.2050\times10^7\,\mathrm m$, $\kappa_n\simeq0.5114997364$, and $q_n\simeq0.3580498155$.

Starting from the spherical reference configuration $a_1=a_2=a_3=R_0$, the affine model describes tidal deformation through three semi-axes and two angular degrees of freedom: the figure angle $\psi$, which specifies the orientation of the ellipsoid relative to the parallel-transported frame, and the internal fluid angle $\lambda$, which parametrizes the motion of fluid elements within the ellipsoid. Their proper time rates are
\begin{equation}
 \Omega=\frac{d\psi}{d\tau},\qquad
 \Lambda=\frac{d\lambda}{d\tau}.
 \label{angles}
\end{equation}
The figure rate $\Omega$ is thus distinct from the coordinate orbital frequency $\Omega_{\rm orb}$. We define the stellar principal axis frame by an orthonormal spatial basis $\boldsymbol e_i$ aligned with the ellipsoid semi-axes $a_i$. Relative to the parallel-transported spatial basis $\boldsymbol\lambda_{\hat i}$, the stellar basis is $\boldsymbol e_i=T(\psi)_{ij}\boldsymbol\lambda_{\hat j}$: it rotates in the radial--tangential plane while its polar leg remains unchanged. To couple the stellar axes to the curvature calculated in Sec.~\ref{Sec.3 A}, we therefore express the electric tidal tensor of Eq.~\eqref{tidal tensor} in this stellar principal axis frame:
\begin{equation}
 c=T(\psi)\mathcal ET(\psi)^T,\qquad
 T(\psi)=\begin{pmatrix}
 \cos\psi&0&\sin\psi\\0&1&0\\-\sin\psi&0&\cos\psi
 \end{pmatrix}.
 \label{eq:affine_rotation}
\end{equation}
The leading quadrupolar tidal potential follows from the local Fermi normal expansion along the WD center of mass geodesic. Expressed in the stellar principal axis coordinates $X^i$, measured relative to the WD center, it is
\begin{equation}
 \Phi_{\rm tide}=\frac12 c_{ij}X^iX^j.
 \label{eq:fermi_tidal_energy}
\end{equation}
We retain this quadrupolar coupling, assuming that the stellar size is small compared with the curvature variation scale and neglecting higher spatial curvature gradients and velocity dependent Fermi terms \cite{ManasseMisner1963,IshiiShibataMino2005}.

The affine Lagrangian separates the stellar contribution from the external tidal interaction \cite{CarterLuminet1985,Lai:1994hf,Wiggins_2000}:
\begin{equation}
 \mathcal L=\mathcal L_{\rm B}+\mathcal L_{\rm T},\qquad
 \mathcal L_{\rm B}=T_{\rm aff}-U-W_{\rm self}.
 \label{Lagrange}
\end{equation}
Here $\mathcal L_{\rm B}$ describes the WD without external tides and consists of the affine kinetic energy $T_{\rm aff}$, internal energy $U$, and Newtonian self gravitational energy $W_{\rm self}$. For a compressible ellipsoid, these terms take the forms
\begin{align}
 T_{\rm aff}&=\frac{\kappa_nm}{10}\biggl[
 \sum_i\left(\frac{da_i}{d\tau}\right)^2
 +(a_1^2+a_3^2)(\Omega^2+\Lambda^2)
 -4a_1a_3\Omega\Lambda\biggr],\notag\\
 U&=\frac{nm^2}{(5-n)R_0}\left(\frac{R_0^3}{a_1a_2a_3}\right)^{1/n},\notag\\
 W_{\rm self}&=-\frac{3m^2}{2(5-n)}
 \int_0^\infty\frac{du}{\sqrt{\prod_i(a_i^2+u)}}.
 \label{eq:affine_internal_energies}
\end{align}
Integrating the quadrupolar tidal potential of Eq.~\eqref{eq:fermi_tidal_energy} over the stellar mass distribution gives the tidal contribution in terms of the second mass moment $I_{ij}$:
\begin{equation}
 \mathcal L_{\rm T}=-\frac12 c_{ij}I_{ij},\qquad
 I_{ij}=\frac{\kappa_nm}{5}a_i^2\delta_{ij},\qquad i,j=1,2,3.
 \label{eq:second_mass_moment}
\end{equation}
The angular degrees of freedom have canonical momenta $J_s=\partial\mathcal L/\partial\Omega$ and $\mathcal C=\partial\mathcal L/\partial\Lambda$, respectively:
\begin{align}
 J_s&=\frac{\kappa_nm}{5}
 [(a_1^2+a_3^2)\Omega-2a_1a_3\Lambda],\notag\\
 \mathcal C&=\frac{\kappa_nm}{5}
 [(a_1^2+a_3^2)\Lambda-2a_1a_3\Omega].
 \label{eq:affine_momenta}
\end{align}
The Lagrangian has no explicit dependence on the internal fluid angle $\lambda$. Consequently, its conjugate momentum $\mathcal C$, corresponding to the fluid circulation, is conserved. By contrast, the stellar angular momentum $J_s$ can evolve under an external tidal torque.

To express the Euler--Lagrange equations in a form suitable for numerical equilibrium calculations, we scale the semi-axes by the undeformed WD radius $R_0$ and proper time by the WD dynamical timescale $t_*$. In geometrized units, the dimensionless quantities are
\begin{align}
 t_*&=\sqrt{R_0^3/m},&\bar\tau&=\tau/t_*,&\bar a_i&=a_i/R_0,\notag\\
 \bar\Omega&=t_*\Omega,&\bar\Lambda&=t_*\Lambda,
 &\bar c_{ij}&=t_*^2c_{ij},\notag\\
 \bar J_s&=\frac{J_s}{\sqrt{m^3R_0}},
 &\bar{\mathcal C}&=\frac{\mathcal C}{\sqrt{m^3R_0}}.
 \label{eq:affine_dimensionless}
\end{align}
We also introduce the dimensionless volume factor $\bar V=\bar a_1\bar a_2\bar a_3$. With overdots denoting derivatives with respect to $\bar\tau$, the Euler--Lagrange equations in these dimensionless variables take the form \cite{Lai:1994hf,Wiggins_2000}:
\begin{align}
 \ddot{\bar a}_1&=\bar a_1(\bar\Omega^2+\bar\Lambda^2)
 -2\bar a_3\bar\Omega\bar\Lambda
 -\frac{3\bar a_1\bar A_1}{2q_n}
 +\frac{1}{q_n\bar a_1\bar V^{1/n}}
 -\bar c_{11}\bar a_1,\label{eq of motion 1}\\
 \ddot{\bar a}_2&=-\frac{3\bar a_2\bar A_2}{2q_n}
 +\frac{1}{q_n\bar a_2\bar V^{1/n}}
 -\bar c_{22}\bar a_2,\label{eq of motion 2}\\
 \ddot{\bar a}_3&=\bar a_3(\bar\Omega^2+\bar\Lambda^2)
 -2\bar a_1\bar\Omega\bar\Lambda
 -\frac{3\bar a_3\bar A_3}{2q_n}
 +\frac{1}{q_n\bar a_3\bar V^{1/n}}
 -\bar c_{33}\bar a_3,\label{eq of motion 3}\\
 \dot{\bar J}_s&=\frac{\kappa_n}{5}\bar c_{13}
 (\bar a_3^2-\bar a_1^2),\label{eq of motion 4}\\
 \dot{\bar{\mathcal C}}&=0.\label{eq of motion 5}
\end{align}
The coefficients $\bar A_i$ appearing in the semi-axis equations are the ellipsoidal shape integrals,
\begin{equation}
 \bar A_i=\int_0^\infty
 \frac{du}{(\bar a_i^2+u)\sqrt{\prod_j(\bar a_j^2+u)}}.
 \label{eq:shape_integrals}
\end{equation}
For numerical black hole backgrounds, the tidal coefficients $\bar c_{ij}$ and the frame rate are converted to the WD units defined above; the conversion from the black hole solver's length scale is given in Appendix~\ref{app:WD_units}.

The full affine equations describe stellar deformation that changes with time. To construct a quasi-equilibrium sequence instead, we assume that orbital evolution is slow compared with the WD dynamical timescale and seek constant semi-axes at each orbital radius. We select the irrotational branch by setting the conserved circulation $\mathcal C=0$. For aligned quasiequilibria, the WD principal axes follow the radial, polar, and azimuthal directions of the orbit-adapted tetrad. Since the figure angle $\psi$ is measured relative to the parallel-transported frame, this alignment requires $\psi=\xi$, so that $\Omega=\omega_{\rm fr}$ and $\bar\Omega=t_*\omega_{\rm fr}$. The resulting conditions are
\begin{equation}
 \dot{\bar a}_i=\ddot{\bar a}_i=0\quad(i=1,2,3),\qquad
 \psi=\xi,\qquad\bar\Omega=\frac{d\xi}{d\bar\tau},\qquad
 \bar{\mathcal C}=0.
 \label{eq:quasiequilibrium_conditions}
\end{equation}
In particular, setting $\bar{\mathcal C}=0$ in Eq.~\eqref{eq:affine_momenta} determines the internal fluid rate:
\begin{equation}
 \bar\Lambda=\frac{2\bar a_1\bar a_3}{\bar a_1^2+\bar a_3^2}\bar\Omega.
 \label{eq:zero_circulation}
\end{equation}
Zero circulation does not imply $\Lambda=0$: the internal fluid rate adjusts as the ellipsoidal figure rotates while the flow remains irrotational. Moreover, $\psi=\xi$ aligns the stellar axes with the orbit-adapted tidal principal directions identified in Sec.~\ref{Sec.3 A}. In this configuration $c_{13}=0$, so the tidal torque in Eq.~\eqref{eq of motion 4} vanishes. With the conserved circulation fixed, the quasi-equilibrium problem thus reduces to three static semi-axis equations at each orbital radius.

\subsection{Critical quasi-equilibrium configurations}
The physical significance of a limiting tidal equilibrium is most
clearly seen by following the WD equilibrium shapes inward along a
sequence of circular orbits. Far from the black hole, the tidal field is weak
and the star is nearly spherical. As the orbital radius decreases, the
ellipsoid can become more elongated in the radial than in the tangential
direction, so the ratio $x\equiv\bar a_3/\bar a_1$ decreases.
The sequence connected to this weak tide configuration may reach a
limiting orbital radius $r_{\rm fold}$ \cite{Shibata1996}. There
the branch followed inward from large radius reaches a turning point
in the orbital radius. Under the usual nondegeneracy conditions, this
is a saddle-node (fold) point of the static equilibrium equations
\cite{Wiggins_2000,Kuznetsov2004}.

We introduce shape coordinates $\boldsymbol y=(\bar a_1,\bar a_2,x)$
with $x>0$. The third semi-axis and dimensionless volume factor are
\begin{equation}
 \bar a_3=x\bar a_1,\qquad \bar V=\bar a_1^2\bar a_2x.
 \label{eq:shape_coordinates}
\end{equation}
For each circular orbit radius $r$, define the static residual
$\boldsymbol F(\boldsymbol y,r)=(F_1,F_2,F_3)^T$ by the right hand
sides of the three semi-axis equations
\eqref{eq of motion 1}--\eqref{eq of motion 3}. In forming this residual,
we impose the aligned zero circulation conditions, eliminate
$\bar\Lambda$ using Eq.~\eqref{eq:zero_circulation}, and apply
Eq.~\eqref{eq:shape_coordinates}. The tidal tensor and the frame
rotation rate are evaluated at the selected circular orbit with the
unit conversions of Appendix~\ref{app:WD_units}.

At fixed orbital radius, form the shape Jacobian
\begin{equation}
 J_y=\left.\frac{\partial\boldsymbol F}{\partial\boldsymbol y}\right|_r.
 \label{eq:shape_jacobian}
\end{equation}
A regular equilibrium has nonsingular $J_y$, whereas an ordinary
fold has $\boldsymbol F=0$ and $\operatorname{rank}J_y=2$.
The additional transversality and quadratic nondegeneracy conditions
are given in Appendix~\ref{app:fold}. A local fold is not by itself
sufficient to identify a physical tidal limit. We additionally require:
\begin{itemize}
  \item[$\bullet$] The equilibrium belongs to the sequence connected
  continuously to the nearly spherical WD in the weak tide regime.
  \item[$\bullet$] All three WD semi-axes are positive and finite.
  \item[$\bullet$] The WD center follows a timelike circular orbit
  outside the black hole horizon.
\end{itemize}
We also compare the fold radius with the ISCO. A fold inside the ISCO is a valid local mathematical
fold but need not be reached during an inspiral along radially stable
circular orbits. This distinction between loss of stellar equilibrium and loss of orbital stability guides the comparisons of the two critical radii in Sec.~\ref{sec:numerical_results}.

\section{Numerical results}\label{sec:numerical_results}


We evaluate the static tidal equilibria for a WD with
$m=0.40M_\odot$, polytropic index $n=3/2$, and $\mu_e=2$,
orbiting a black hole of mass $2\times10^5M_\odot$.
The Lane--Emden stellar model gives
$\kappa_{3/2}\simeq0.5114997364$ and
$\bar R_0\simeq3.2385443224$ (see in Appendix~\ref{app:WD_units}).
We keep this stellar model and the aligned, zero-circulation
affine equilibrium sequence fixed throughout the comparison.
The tidal tensors are evaluated using numerical EsGB black hole
backgrounds at spectral resolution $(N_x,N_\theta)=(72,16)$
for rotating configurations. For static configurations ($\chi=0$),
the resolution is $(96,0)$.

Each EsGB fold is compared with a Kerr fold at the same ADM mass,
angular momentum, stellar model, and orbital branch. This matching
uses asymptotic charges rather than equal quasi-isotropic coordinate
radii. The orbital frequency provides a common measure of the
critical configuration. We define its dimensionless magnitude
and the relative EsGB--Kerr correction by
\begin{equation}
 w_{\rm fold}=M_{\rm ADM}\bigl|\Omega_{\rm orb}(r_{\rm fold})\bigr|,\qquad
 \delta_\Omega=\frac{w_{\rm fold}^{\rm EsGB}-w_{\rm fold}^{\rm Kerr}}
 {w_{\rm fold}^{\rm Kerr}}.
 \label{eq:critical_frequency_definition}
\end{equation}
The magnitude $w_{\rm fold}$ permits a direct comparison of
prograde and retrograde orbits, while the signed
$\Omega_{\rm orb}$ enters their geodesic and tidal equations. The relative correction $\delta_\Omega$ then measures how the theory-dependent tidal geometry shifts the fold frequency after the mass and spin dependence of the Kerr reference has been removed.

\subsection{Coupling dependence at fixed spin}\label{subsec:results_fixed_spin}

\begin{figure*}[t]
 \centering
 \begin{minipage}[t]{0.49\linewidth}
  \centering
  \includegraphics[width=\linewidth]{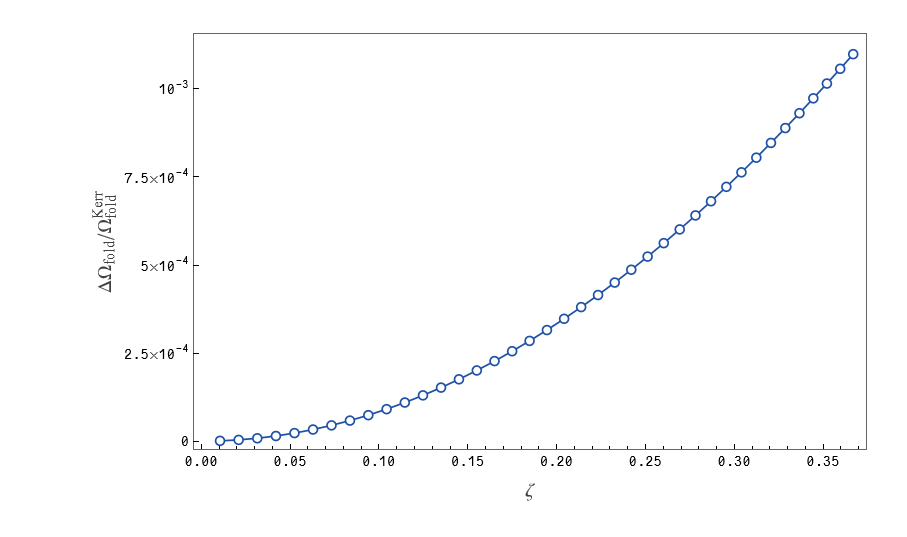}
  \par\small (a)
 \end{minipage}\hfill%
 \begin{minipage}[t]{0.49\linewidth}
  \centering
  \includegraphics[width=\linewidth]{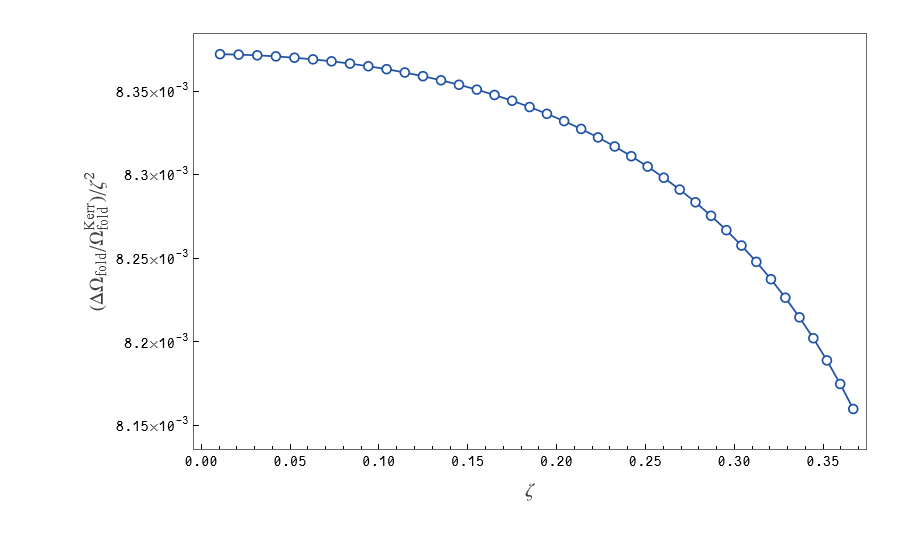}
  \par\small (b)
 \end{minipage}
 \caption{Fixed-spin ($\chi=0.4$) EsGB--Kerr comparison:
 (a) relative shift in fold frequency $\delta_\Omega$;
 (b) $\delta_\Omega/\zeta^2$. Each point compares the
 critical frequencies on EsGB and Kerr backgrounds matched in
 ADM mass and spin.}
 \label{fig:results_fixed_delta}
\end{figure*}

With the matching prescription established, Fig.~\ref{fig:results_fixed_delta}(a) shows how the relative fold frequency correction varies with coupling at fixed spin $\chi=0.4$.
The correction is positive throughout the plotted sequence and grows
with $\zeta$, reaching order $10^{-3}$ at the upper end of the
sampled coupling range. For the representative parameters $(r_H,\chi,\alpha)=(0.1,0.4,0.002)$, the matched fold frequencies and their difference are
\begin{equation}
  w_{\rm fold}^{\rm EsGB}\simeq0.05598041,\quad
  w_{\rm fold}^{\rm Kerr}\simeq0.05597959,\quad
  \delta_\Omega\simeq1.4733\times10^{-5},\quad
  \zeta\simeq0.04195.
 \label{eq:results_benchmark}
\end{equation}
The representative configuration fixes the scale of the correction at one coupling; its functional dependence is clearer from the full sequence. Figure~\ref{fig:results_fixed_delta}(b) therefore shows $\delta_\Omega/\zeta^2$ to test the leading coupling behavior. This ratio stays near
$8\times10^{-3}$ and decreases gradually as the coupling grows,
whereas $\delta_\Omega$ changes by orders of magnitude.
The comparison therefore supports an approximately quadratic
correction at weak coupling,
\begin{equation}
 \delta_\Omega(\zeta,\chi=0.4)
 =C_2(\chi=0.4)\,\zeta^2+\text{higher-order terms}.
 \label{eq:results_weak_coupling}
\end{equation}
Here $C_2(\chi)$ denotes the coefficient of the leading quadratic term in the weak-coupling expansion at fixed spin. In the linear theory, the scalar field is sourced at first
order in the coupling, whereas its leading backreaction on the
metric enters at second order. The orbital motion, the electric
part of the Riemann tensor in the stellar frame, and the frame
rotation rate therefore acquire leading geometric corrections of
this order. These quantities enter the affine force balance, so a
nearly quadratic shift of the fold frequency is the expected
signature of the scalar modification of the tidal geometry. The
behavior in Fig.~\ref{fig:results_fixed_delta}(b) is consistent
with this mechanism.

\begin{figure}[tbp]
 \centering
 \includegraphics[width=0.7\linewidth]{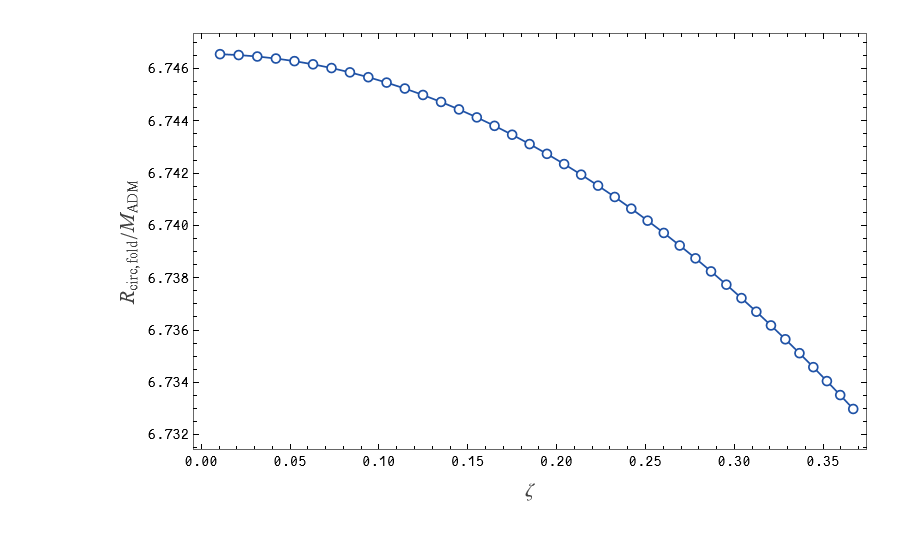}
 \caption{Equatorial circumferential radius at the EsGB static
 fold, $R_{\rm circ}/M_{\rm ADM}$, for the fixed $\chi=0.4$
 coupling sequence. The radius is defined from the azimuthal
 metric component, Eq.~\eqref{eq:results_circ_radius},
 so the plotted quantity has a geometric interpretation beyond the
 chosen radial coordinate.}
 \label{fig:results_circ_radius}
\end{figure}

The fold location provides a complementary geometric diagnostic.
Figure~\ref{fig:results_circ_radius} shows the circumferential
radius at the EsGB fold, normalized by the ADM mass and defined on
the equatorial plane by
\begin{equation}
 R_{\rm circ}(r)\equiv
 \sqrt{g_{\varphi\varphi}(r,\pi/2)}.
 \label{eq:results_circ_radius}
\end{equation}
The normalized circumferential radius decreases smoothly from
approximately $6.747$ to $6.733$ as $\zeta$ increases across this
sequence, while the matched fold frequency correction is positive.
The two observables thus locate the limiting stellar equilibrium at
a smaller circumferential radius and a higher orbital frequency as
the EsGB coupling grows. This connects the frequency effect to a
change in the geometry of the tidal equilibrium itself: the balance
between the WD's internal forces and the locally projected curvature
is reached at a different orbital location. The use of the
circumferential radius makes this geometric trend independent of the
particular radial coordinate used to label the orbit. Having established the coupling response at one spin, we next examine how rotation changes both the limiting frequency and its EsGB correction.

\subsection{Dependence on black hole spin}\label{subsec:results_spin}

\begin{figure*}[t]
 \centering
 \begin{minipage}[t]{0.49\linewidth}
  \centering
  \includegraphics[width=\linewidth]{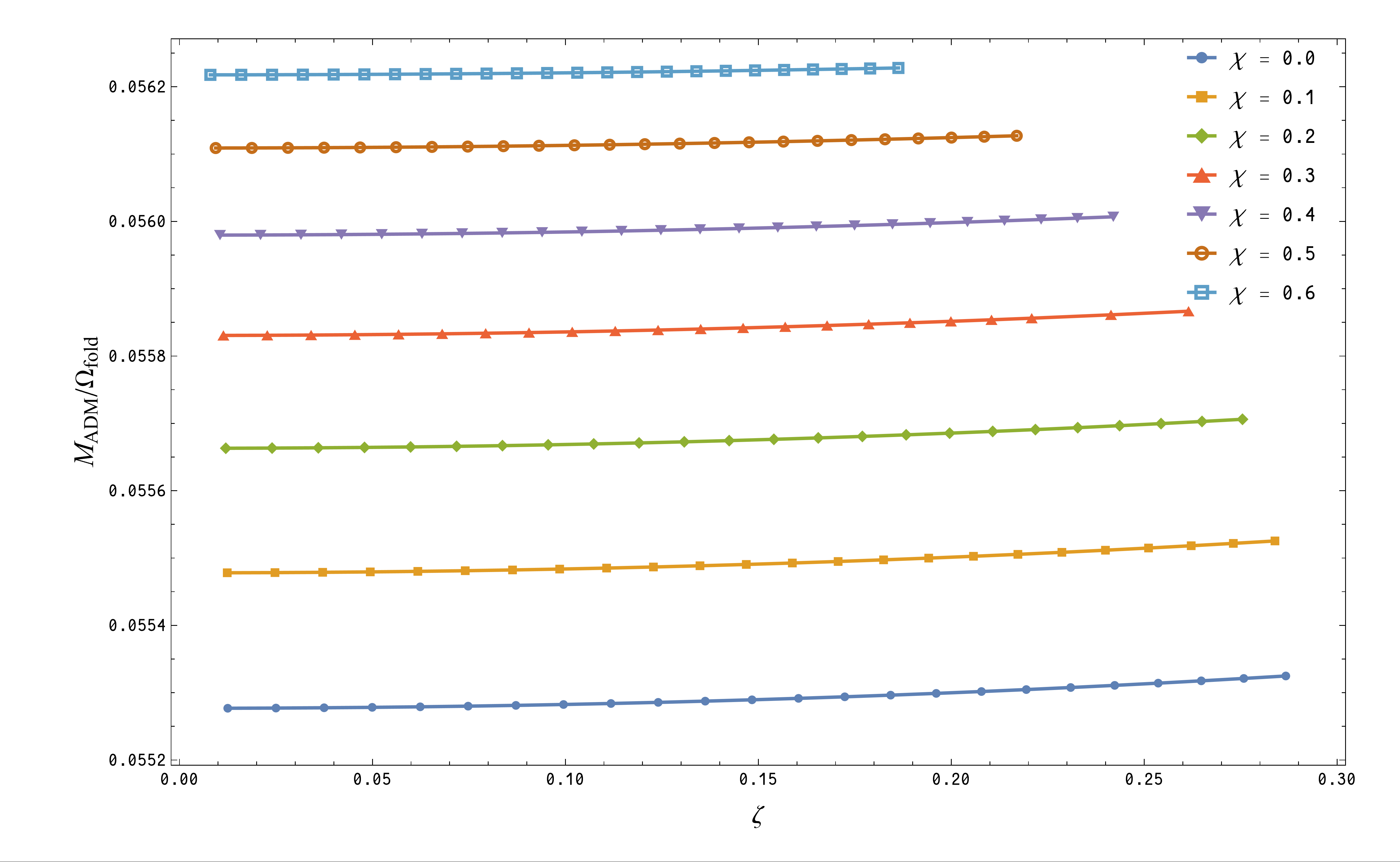}
  \par\small (a)
 \end{minipage}\hfill%
 \begin{minipage}[t]{0.49\linewidth}
  \centering
  \includegraphics[width=\linewidth]{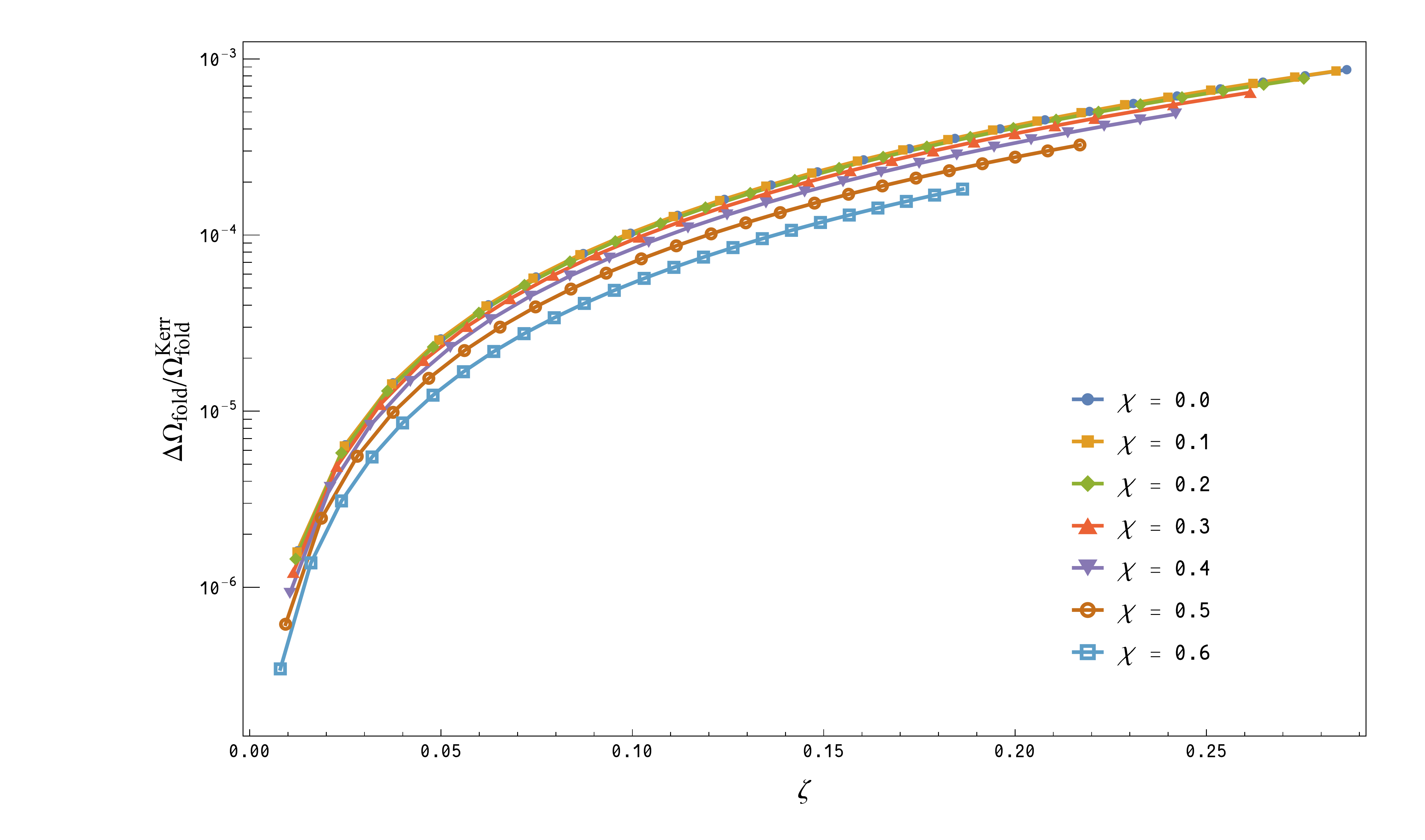}
  \par\small (b)
 \end{minipage}
 \caption{Spin dependence of matched static folds:
 (a) critical orbital frequency normalized by mass;
 (b) fractional EsGB--Kerr shift for seven spin branches.
 Each spin branch covers its sampled coupling interval.}
 \label{fig:results_spin_w}
\end{figure*}

Figure~\ref{fig:results_spin_w}(a) shows the dimensionless
EsGB fold frequency for seven spin branches. At fixed spin,
$w_{\rm fold}$ increases gently with coupling; the separation between
spin branches is considerably larger than the variation along any
one branch. Black hole rotation therefore has a pronounced influence
on the orbital frequency at which the WD reaches its limiting
equilibrium. Figure~\ref{fig:results_spin_w}(b) isolates the
additional EsGB contribution through the matched relative
correction $\delta_\Omega$.

The matched correction remains positive across the displayed
branches and reaches order $10^{-3}$ in the strongest coupling
samples. In the overlapping coupling ranges, the relative correction
is generally smaller for the more rapidly rotating backgrounds.
Rotation affects both the circular geodesic frequency and the
curvature projected into the WD's local frame; it therefore changes
how the scalar modification enters the force balance at the fold.
The different spin branches show that the shift cannot be described
by coupling alone. Matching the Kerr mass and spin isolates the
additional effect of the EsGB geometry, while the residual spin
dependence reveals its interaction with black hole rotation. A shift in the fold frequency, however, does not by itself determine whether the WD can approach that configuration along a radially stable circular orbit. This requires a separate comparison with the ISCO.

\FloatBarrier

\subsection{Prograde and retrograde folds relative to the ISCO}
\label{subsec:results_fold_isco}

\begin{figure}[H]
 \centering
 \begin{minipage}[t]{0.485\linewidth}
  \centering
  \includegraphics[width=\linewidth]{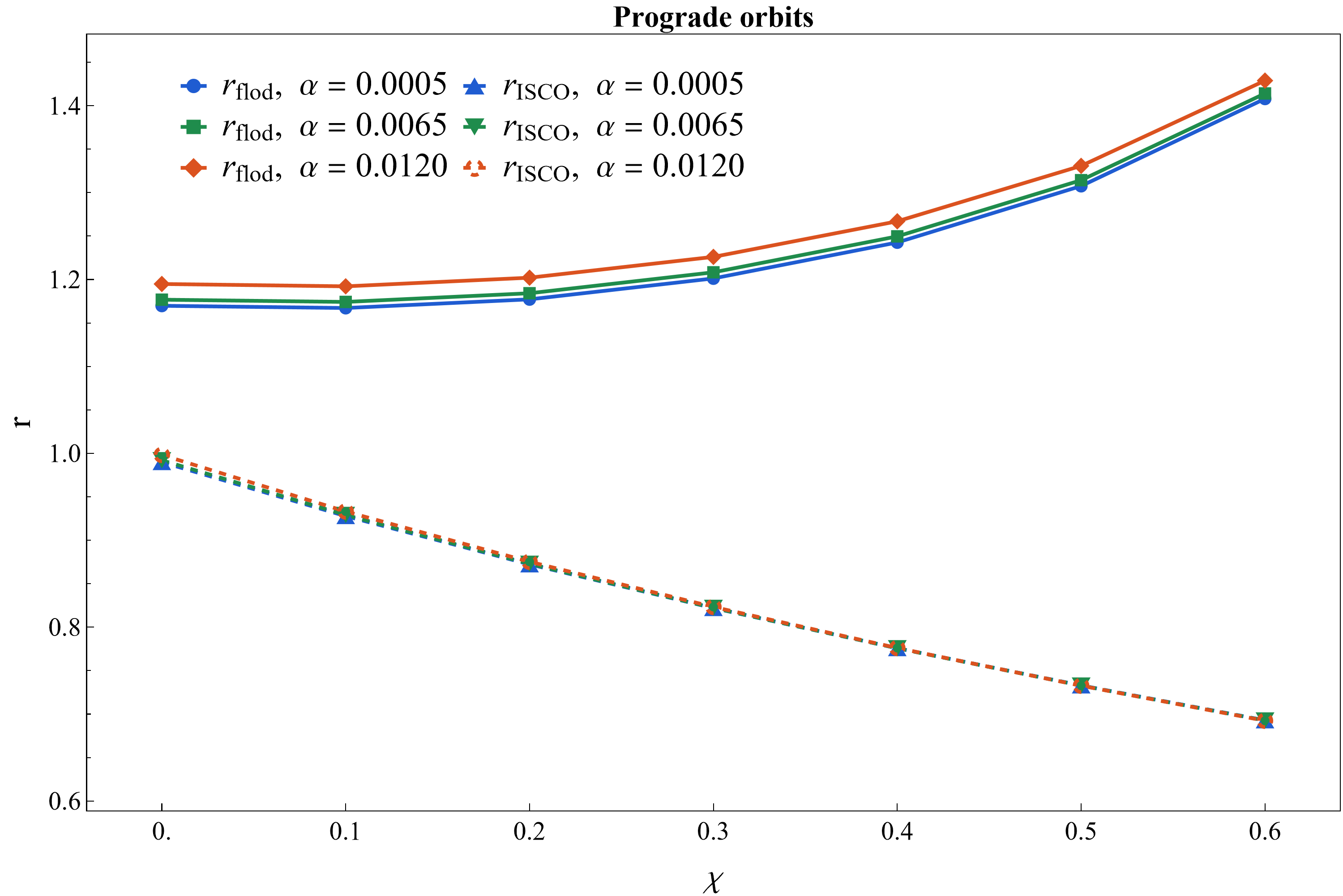}
  \par\small (a)
 \end{minipage}\hfill%
 \begin{minipage}[t]{0.485\linewidth}
  \centering
  \includegraphics[width=\linewidth]{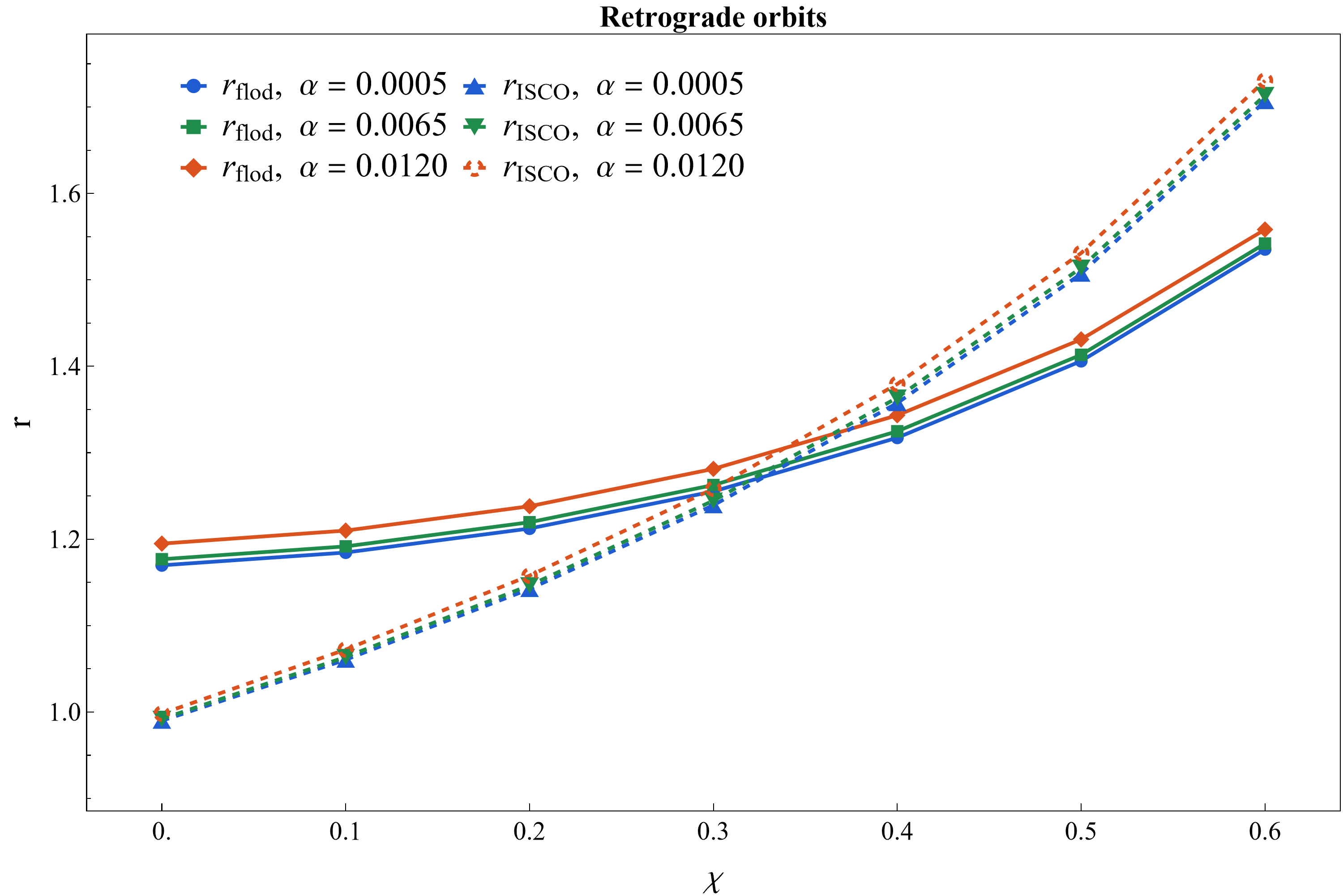}
  \par\small (b)
 \end{minipage}
 \caption{Tidal fold and ISCO radii for circular orbits on the same
 numerical EsGB backgrounds: (a) prograde and (b) retrograde.
 Solid curves show $r_{\rm fold}$ and dashed curves show $r_{\rm ISCO}$.
 The curves correspond to $\alpha=0.0005$, $0.0065$, and $0.0120$. All sampled prograde
 folds lie outside the ISCO; for retrograde orbits the ordering reverses
 between $\chi=0.3$ and $0.4$.}
 \label{fig:results_orbit_isco}
\end{figure}

\begin{figure}[H]
 \centering
 \includegraphics[width=0.7\linewidth]{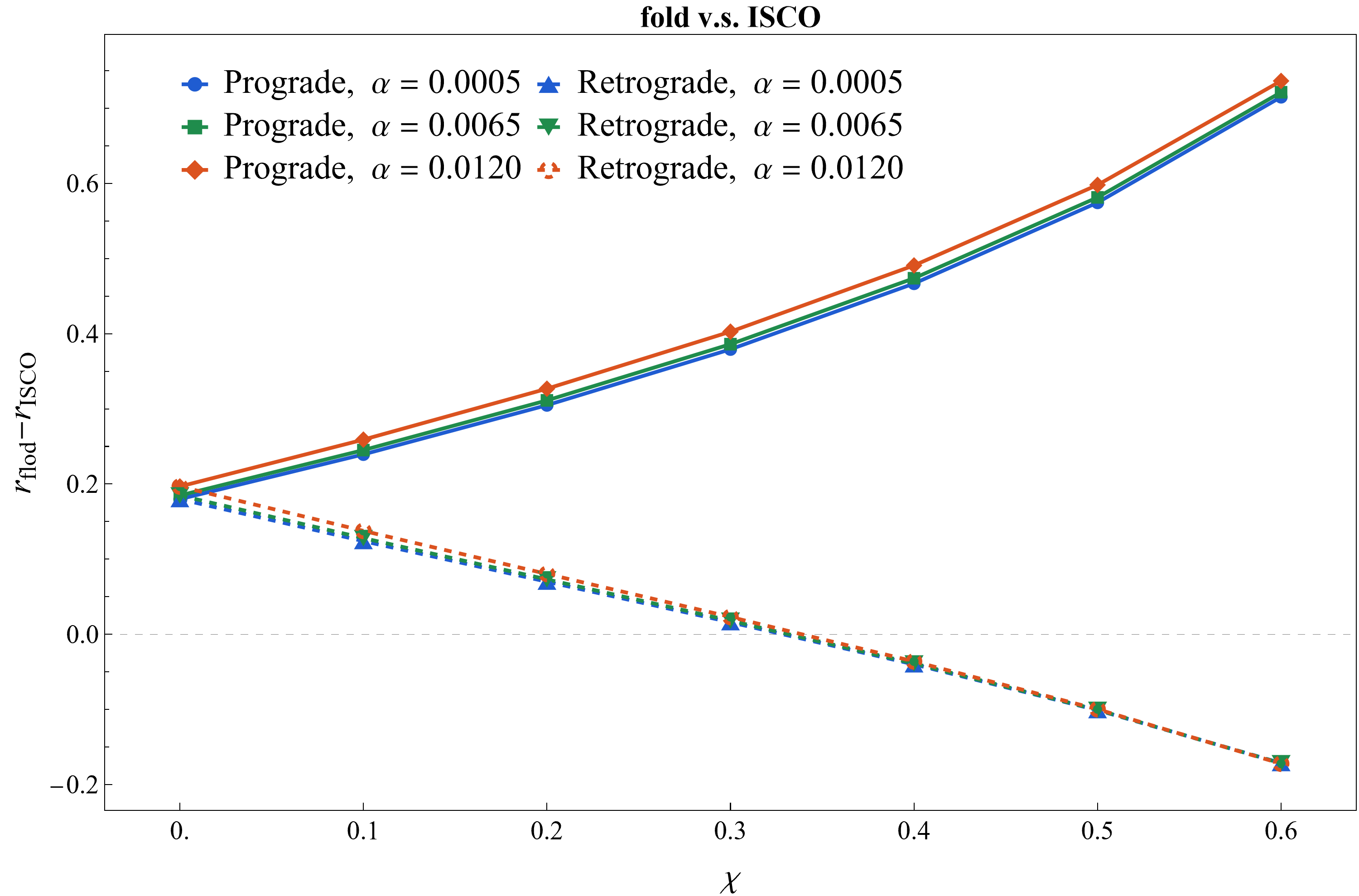}
 \caption{Difference between the tidal fold and ISCO coordinate
 radii, $r_{\rm fold}-r_{\rm ISCO}$,
 on the same EsGB background, plotted separately for prograde
 (solid) and retrograde (dashed) motion. Positive (negative) values
 correspond to a fold outside (inside) the ISCO. The horizontal line
 marks zero. The retrograde sign change is resolved only between the
 sampled values $\chi=0.3$ and $0.4$.}
 \label{fig:results_fold_minus_isco}
\end{figure}

To determine which limiting configuration is encountered along a sequence of stable circular orbits, we compare the tidal fold with the ISCO. These are distinct physical thresholds: the fold
marks the loss of the selected stellar equilibrium branch, whereas
the ISCO marks the loss of radial stability of the center of mass
orbit. Their relative positions determine which limit is encountered
first along an inward sequence of stable circular orbits. We evaluate both radii on the same EsGB background and for the same orbital orientation. Figures~\ref{fig:results_orbit_isco} and \ref{fig:results_fold_minus_isco} show the three couplings $\alpha=0.0005$, $0.0065$, and $0.0120$.

In Fig.~\ref{fig:results_orbit_isco}(a), every sampled prograde
fold lies outside the ISCO over $\chi=0,0.1,\ldots,0.6$.
As spin increases, the prograde ISCO moves inward while the fold
radius varies much less and rises slightly at higher spin.
The separation therefore grows. Along the sampled prograde
sequences, the WD reaches its static tidal equilibrium limit while
its circular center of mass orbit remains radially stable. In this
sense the tidal equilibrium, rather than orbital instability, sets
the first limiting radius for these configurations.

For retrograde motion, Fig.~\ref{fig:results_orbit_isco}(b)
shows that both radii grow with spin, with the ISCO moving outward
more rapidly. Their ordering reverses between $\chi=0.3$ and
$0.4$. At zero spin the two orbital orientations coincide; at
finite spin their different response to rotation produces distinct
locations of the limiting equilibrium. Beyond the crossing, the ISCO lies farther out than the
retrograde static fold. An inward sequence of stable retrograde
circular orbits therefore meets the orbital stability limit before
the limiting affine configuration.

The separation is shown directly in
Fig.~\ref{fig:results_fold_minus_isco} using the radius difference
$r_{\rm fold}-r_{\rm ISCO}$. A positive value means that the fold
is outside the ISCO; a negative value means that the ISCO is
encountered first on an inward stable circular orbit sequence.
The positive difference for every prograde sample confirms that
the prograde tidal fold lies outside the ISCO. For the retrograde
branch, the difference is positive through $\chi=0.3$ and negative
at the sampled spins $\chi\geq0.4$ for all three couplings.
The change in sign is bracketed by $0.3<\chi<0.4$ and identifies
a change in the ordering of the orbital and tidal equilibrium limits.
In particular, the retrograde static fold at larger spins belongs
to the orbitally unstable circular orbit region.

\FloatBarrier

\begin{figure}[tbp]
 \centering
 \begin{minipage}[t]{0.485\linewidth}
  \centering
  \includegraphics[width=\linewidth]{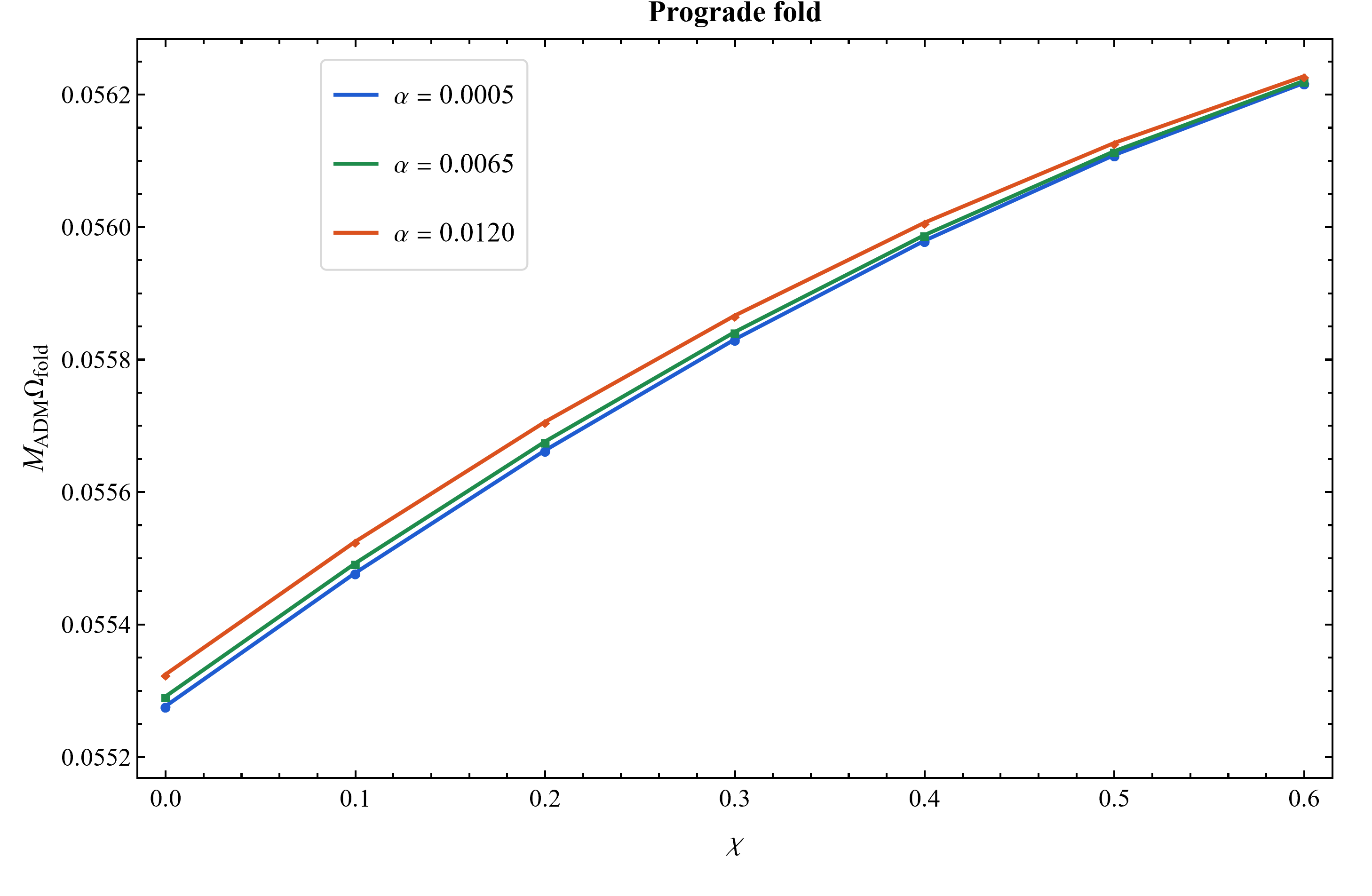}
  \par\small (a)
 \end{minipage}\hfill%
 \begin{minipage}[t]{0.485\linewidth}
  \centering
  \includegraphics[width=\linewidth]{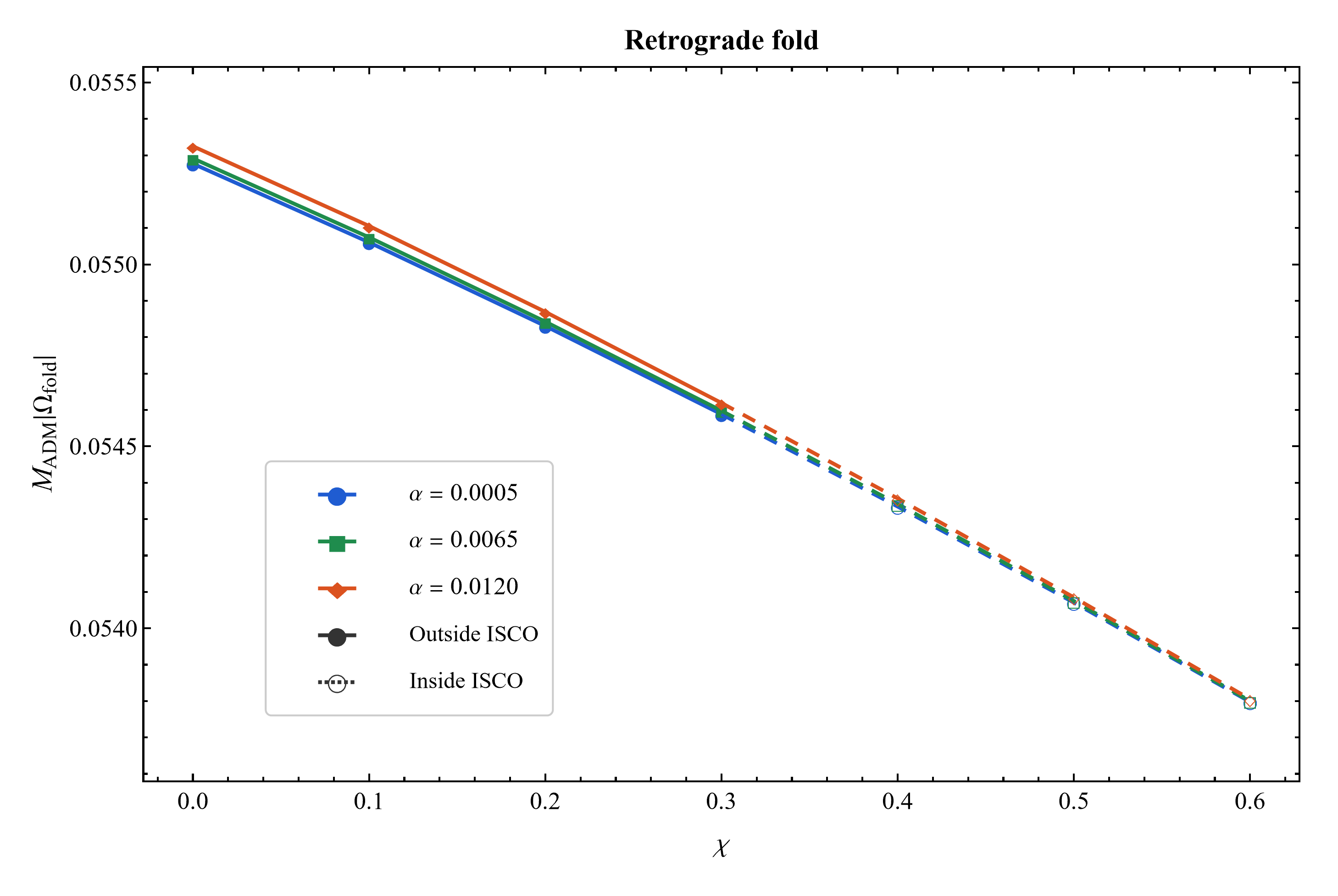}
  \par\small (b)
 \end{minipage}
 \caption{Dimensionless orbital frequency magnitude at the tidal fold,
 $M_{\rm ADM}|\Omega_{\rm fold}|$, versus black hole spin for the couplings $\alpha=0.0005$, $0.0065$, and $0.0120$: (a) prograde and (b) retrograde.
 All displayed prograde folds lie outside the ISCO. In panel (b),
 solid segments and filled markers indicate folds outside the ISCO,
 whereas dashed segments and open markers indicate folds inside it.}
 \label{fig:results_fold_freq}
\end{figure}

The radius comparison establishes which stability limit is encountered first; the fold frequency provides a complementary measure of the orbital conditions at that limit. Figure~\ref{fig:results_fold_freq} shows $w_{\rm fold}=M_{\rm ADM}|\Omega_{\rm orb}(r_{\rm fold})|$ for prograde and retrograde motion in panels (a) and (b), respectively, at the same three couplings. The spin dependence can therefore be read alongside the Fold--ISCO ordering.

For prograde motion, Fig.~\ref{fig:results_fold_freq}(a) shows a
smooth rise of the fold frequency with spin. The change across the
sampled spin range is much larger than the separation of the three
coupling curves. This behavior is consistent with rotation changing
the orbital frequency and curvature environment at the fold; all of
these prograde configurations remain outside the ISCO.

For retrograde motion, Fig.~\ref{fig:results_fold_freq}(b) shows
that the fold frequency decreases with spin, opposite to the
prograde trend. Filled markers and solid curves identify sampled
folds outside the ISCO through $\chi=0.3$; open markers and dashed
curves show the folds inside the ISCO at $\chi\geq0.4$.
The opposite frequency trends, together with the ISCO crossing,
show two linked effects of orbital orientation: frame dragging
changes the orbital frequency at which the WD reaches a static tidal
limit, and it also changes whether that limit is encountered before
or after the loss of orbital stability. The distinction matters when
relating a tidal equilibrium sequence to the orbital evolution of a
WD around a spinning black hole.

\FloatBarrier

The Fold--ISCO comparison identifies which critical equilibria can be approached along radially stable circular orbits. For these configurations, the fold frequency also provides a natural frequency scale for a possible tidal feature in the gravitational wave signal. For nearly circular motion, the dominant quadrupolar harmonic has twice the orbital frequency. Restoring physical constants, the corresponding source frame frequency is
\begin{equation}\label{eq:fold_GW_frequency}
 f_{\rm GW}^{\rm fold}=\frac{c^3}{\pi G M_{\rm ADM}}\,w_{\rm fold}.
\end{equation}
For the $0.40M_\odot$ WD and $2\times10^5M_\odot$ black hole considered here, the calculated fold frequencies span $17.38$--$18.17\,\mathrm{mHz}$, with a representative value of $18.09\,\mathrm{mHz}$. A leading quadrupole estimate at $18\,\mathrm{mHz}$ gives $h_0\simeq1.1\times10^{-21}(10\,\mathrm{Mpc}/D_L)$. This frequency range lies in the bands targeted by LISA \cite{LISAColpi2024} and TianQin \cite{LuoTianQin2016}.

For EsGB and Kerr backgrounds matched in ADM mass and angular momentum, the fractional difference of their fold gravitational wave frequency scales is the same $\delta_\Omega$ obtained from the orbital folds. Thus the coupling and spin dependence of the tidal fold provides a potential way to distinguish the two geometries through gravitational wave observations. If future waveform models identify a measurable tidal feature associated with an accessible fold, observations resolving its displacement relative to Kerr could constrain the EsGB coupling $\zeta$, after accounting for the properties of the source.

\FloatBarrier
\subsection{Numerical convergence and resolvability}
\label{subsec:numerical_convergence}

The tidal tensor depends on derivatives of the spectral metric functions,
making the numerical accuracy of the EsGB black hole backgrounds essential
for determining the tidal fold. We examine five progressively refined
spectral grids, $(N_x,N_\theta)=(40,8),(48,10),(56,12),(64,14)$, and
$(72,16)$, for randomly sampled rotating configurations from our parameter
scan. Here $N_x$ and $N_\theta$ denote the radial and angular resolutions.
For the static resolvability test, we use the radial grids $(80,0)$,
$(88,0)$, and $(96,0)$.

\begin{figure}[H]
 \centering
 \begin{minipage}[t]{0.485\linewidth}
  \centering
  \includegraphics[width=\linewidth]{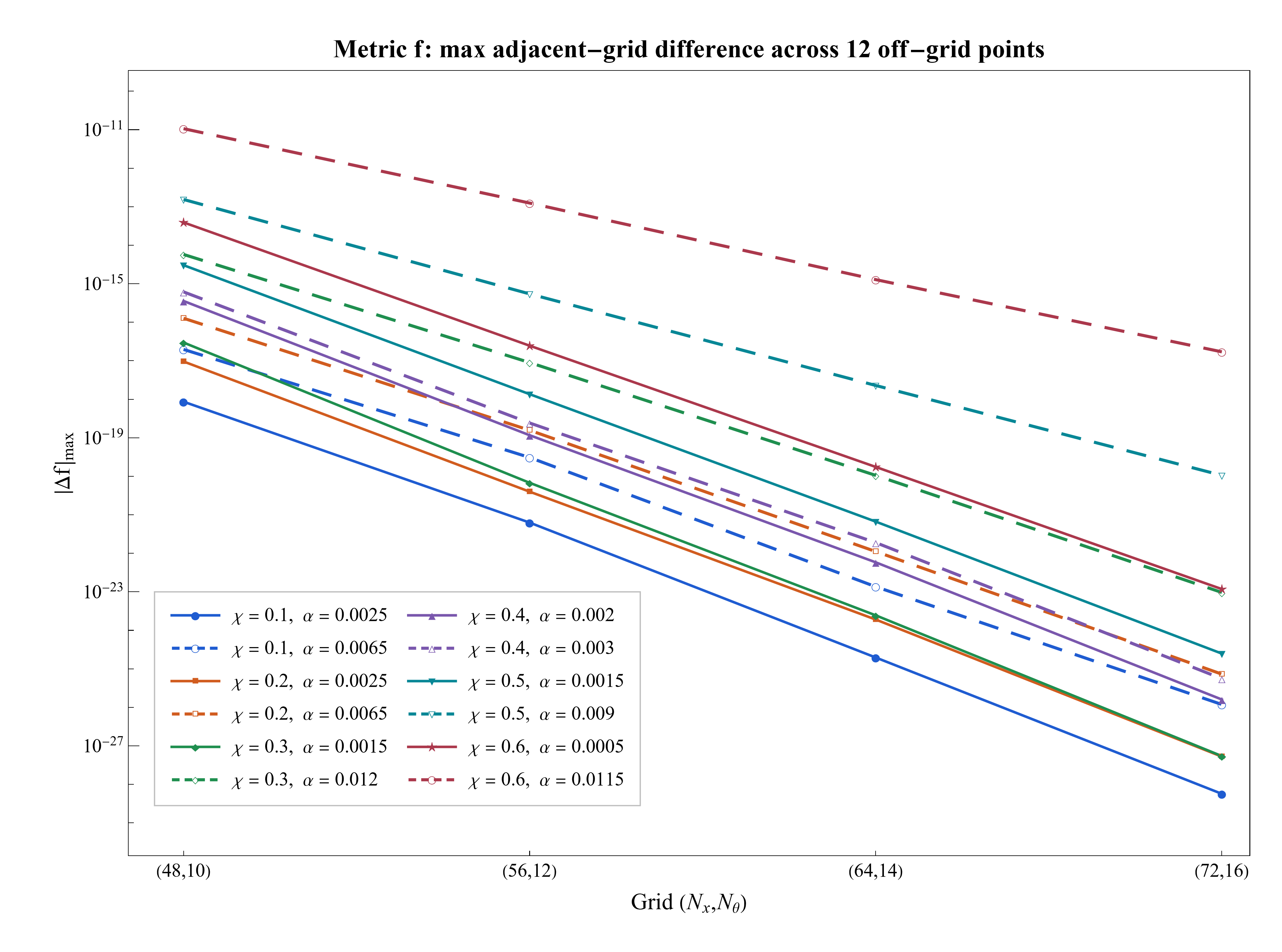}
  \par\small (a)
 \end{minipage}\hfill%
 \begin{minipage}[t]{0.485\linewidth}
  \centering
  \includegraphics[width=\linewidth]{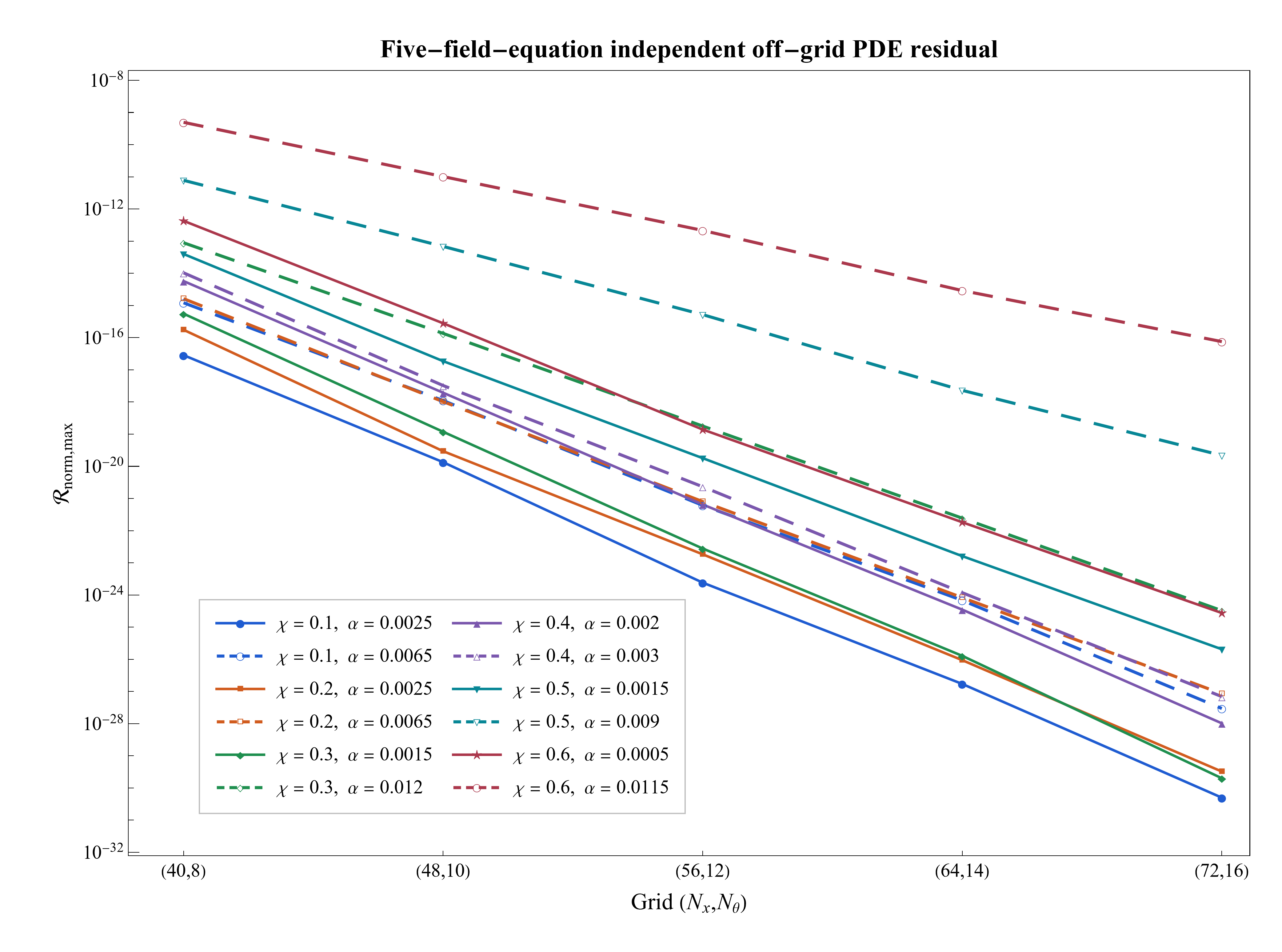}
  \par\small (b)
 \end{minipage}
 \caption{Convergence of randomly sampled rotating EsGB black hole
 backgrounds. (a) Maximum adjacent-resolution difference in the metric
 function $f$, evaluated at 12 fixed off-grid locations. (b) Maximum
 normalized residual among the five EsGB field equations, evaluated away
 from the collocation points.}
 \label{fig:convergence_background}
\end{figure}

Figure~\ref{fig:convergence_background}(a) shows the maximum absolute
difference in the metric function $f$ between adjacent resolutions,
evaluated at 12 fixed locations away from the spectral collocation grid.
Figure~\ref{fig:convergence_background}(b) presents the maximum normalized
residual among the five EsGB field equations at off-grid locations. Both
quantities decrease rapidly with increasing resolution, demonstrating the
convergence of the numerical backgrounds and their consistency with the
field equations.

\begin{figure}[H]
 \centering
 \begin{minipage}[t]{0.485\linewidth}
  \centering
  \includegraphics[width=\linewidth]{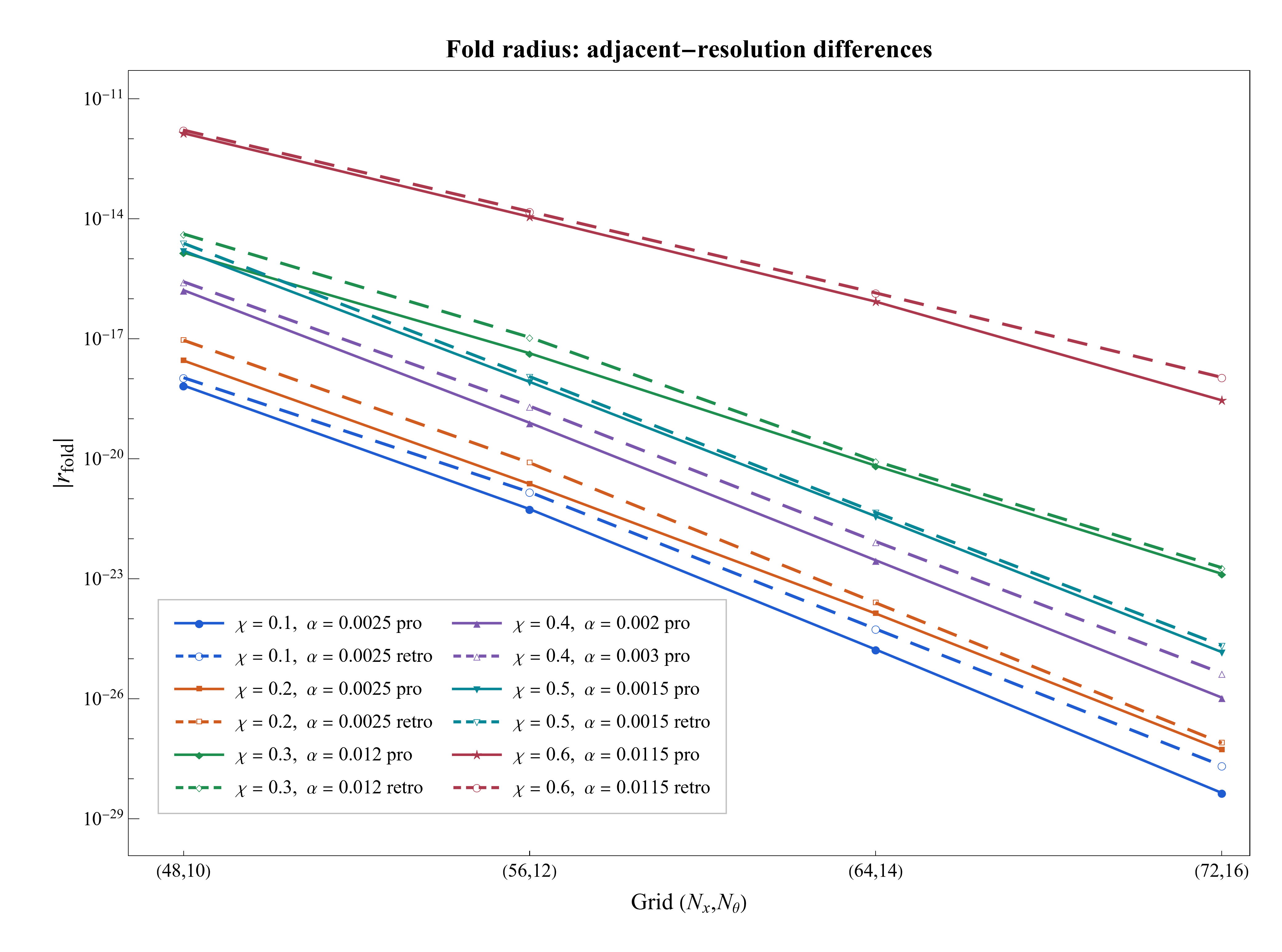}
  \par\small (a)
 \end{minipage}\hfill%
 \begin{minipage}[t]{0.485\linewidth}
  \centering
  \includegraphics[width=\linewidth]{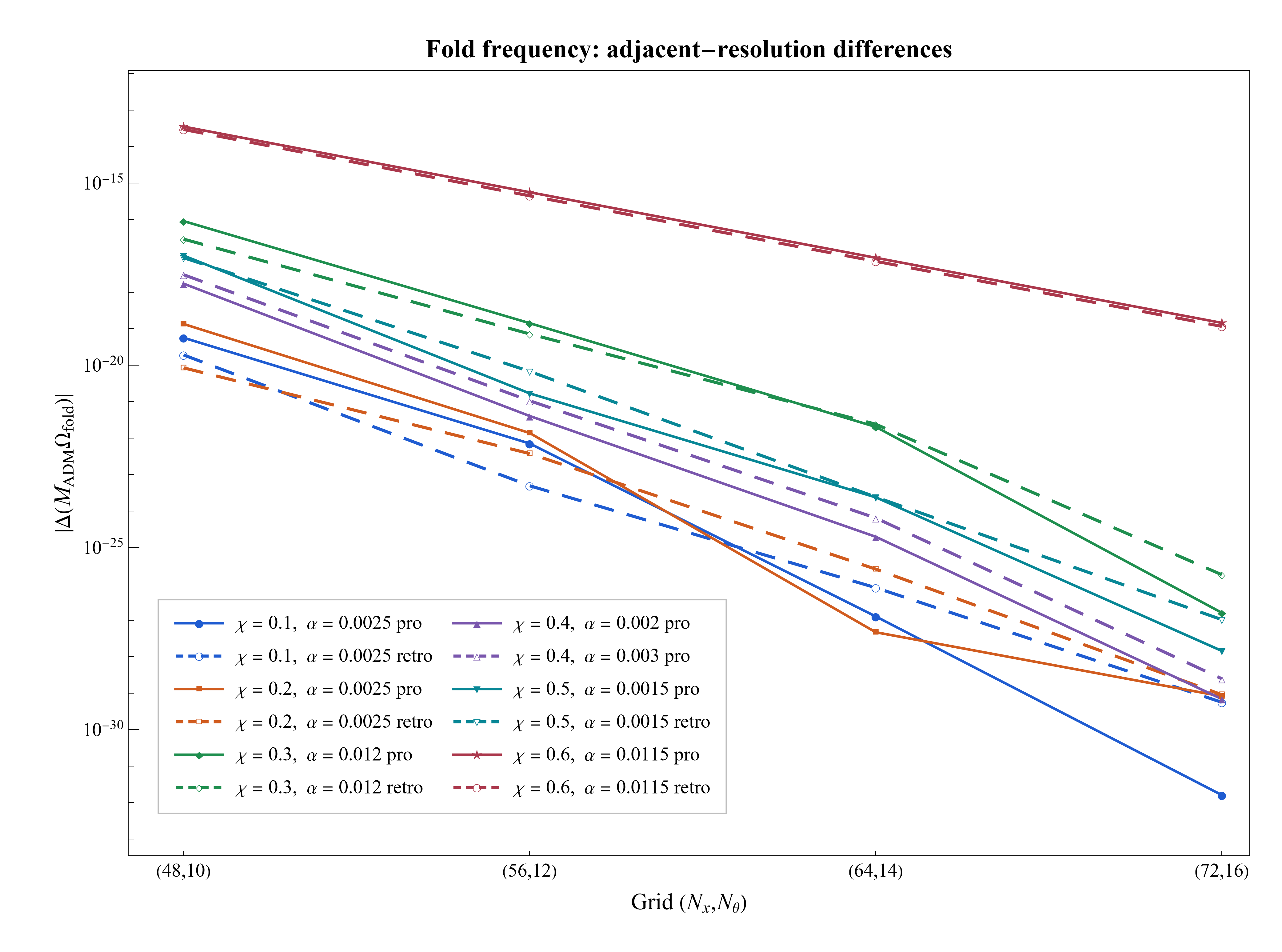}
  \par\small (b)
 \end{minipage}
 \caption{Convergence of the tidal fold quantities for randomly sampled
 rotating configurations. The panels show the absolute adjacent-resolution
 differences in (a) the fold radius $r_{\rm fold}$ and (b) the mass
 normalized frequency $M_{\rm ADM}|\Omega_{\rm fold}|$.}
 \label{fig:convergence_fold}
\end{figure}

We further examine the convergence of the tidal fold quantities.
Figures~\ref{fig:convergence_fold}(a) and \ref{fig:convergence_fold}(b)
show the absolute adjacent-resolution differences in the fold radius
$r_{\rm fold}$ and the mass normalized fold frequency
$M_{\rm ADM}|\Omega_{\rm fold}|$, respectively. Both differences decrease
systematically as the spectral resolution increases, confirming the
numerical stability of the critical affine solutions for the randomly
sampled configurations.

Finally, we assess whether the EsGB--Kerr fold frequency corrections
can be distinguished from the estimated numerical uncertainties. We
define the numerical resolvability ratio as
\begin{equation}
 R_\Omega=\frac{\varepsilon_{\delta_\Omega}}{|\delta_\Omega|},
 \label{eq:resolvability_ratio}
\end{equation}
where $\varepsilon_{\delta_\Omega}$ is the estimated numerical uncertainty
obtained from resolution comparisons, independent high precision fold
solutions, and root sensitivity tests, including the propagation of EsGB
and Kerr frequency uncertainties.

\begin{figure}[H]
 \centering
 \includegraphics[width=0.97\linewidth]{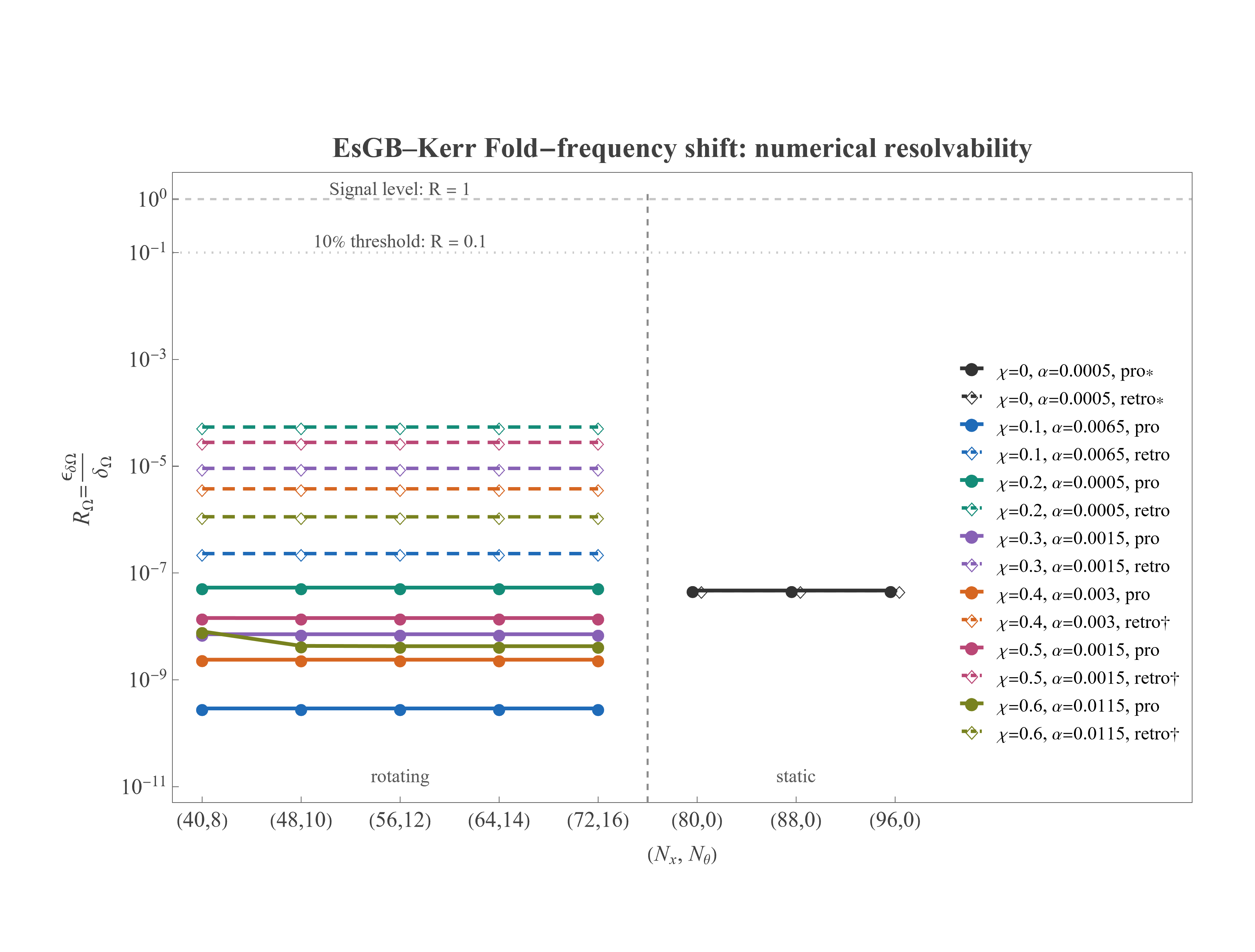}
 \caption{Numerical resolvability of the EsGB--Kerr fold frequency
 corrections for one randomly sampled coupling at each spin. Solid and
 dashed curves represent prograde and retrograde orbits, respectively.
 The static orbital labels share one physical root, and the rightmost grids
 correspond to static backgrounds. Dagger symbols denote mathematical folds
 inside the ISCO. The reference lines indicate $R_\Omega=1$ (signal level)
 and $R_\Omega=0.1$ (10\% threshold).}
 \label{fig:convergence_resolvability}
\end{figure}

For each spin $\chi=0,0.1,\ldots,0.6$, one coupling parameter is randomly
sampled from the parameter scan. Figure~\ref{fig:convergence_resolvability}
shows the corresponding resolvability ratios for prograde and retrograde
orbits. The reference lines $R_\Omega=1$ and $R_\Omega=0.1$ represent the
signal level and the adopted 10\% threshold, respectively. All 14 plotted
sequences, corresponding to 13 independent physical branches, satisfy
$R_\Omega<0.1$. The largest ratio is approximately $5.42\times10^{-5}$,
occurring for the retrograde configuration
$(\chi,\alpha)=(0.2,0.0005)$. The estimated numerical uncertainties are
therefore substantially smaller than the corresponding EsGB--Kerr
frequency corrections for the randomly sampled configurations.

\FloatBarrier

\section{Conclusion}
\label{sec:conclusion}
In this work, we have combined numerical rotating EsGB black hole
backgrounds with a relativistic tidal field and an affine WD model
to determine the limiting tidal equilibria on circular equatorial
orbits. By matching the EsGB and Kerr backgrounds in ADM mass and spin,
we isolate the influence of modified strong-field curvature. The EsGB
fold frequencies exceed their Kerr counterparts in the configurations
investigated. The relative correction grows approximately as $\zeta^2$
at weak coupling and reaches order $10^{-3}$ at the larger couplings
considered. This behavior is consistent with the leading quadratic
metric response to the scalar coupling: changes in orbital motion and
local tidal curvature shift the balance of forces that determines the
critical WD equilibrium.

Spin and orbital orientation also affect the competition between tidal
equilibrium loss and orbital instability. For the prograde configurations
studied, the fold lies outside the ISCO, whereas the ordering reverses
for retrograde motion at higher spins. The two limits therefore need not
be encountered in the same order along inward sequences of stable
circular orbits. This distinction clarifies the role of orbital
stability in interpreting the static tidal threshold.

The predicted EsGB--Kerr differences offer a potential connection to GW
tests of gravity. For the adopted masses, the characteristic fold
frequencies lie in the millihertz band relevant to LISA and TianQin.
The present analysis uses a cold polytropic EOS, Newtonian WD
self-gravity, and quadrupolar tides on a fixed black hole background,
without evolving the dynamical passage through the fold. Future
calculations incorporating orbital evolution, radiation reaction, and
dynamical tidal effects will be needed to identify corresponding
features in GW waveforms and assess whether their EsGB--Kerr shifts
can constrain the coupling $\zeta$.

\section*{Acknowledgments}
This work was supported by the National Key Research and Development Program of China
through the ``Strategic Scientific and Technological Innovation Cooperation'' project
(No.~2025YFE0212600), ``Integrated Electronics Technology for Inertial Sensors''
(No.~2024YFC2207003), and Natural Science Foundation of China under Grant No. 12375056. 
We also acknowledge support from the China--Brazil
Belt and Road Joint Laboratory on Radio Astronomy Technology.

    \FloatBarrier
    \begin{appendices}
    \section{Affine variations and spherical checks}
    \label{app:affine}

    Under a homologous ellipsoidal deformation, let $x_0^i$ denote the fluid coordinates in the undeformed spherical WD and $\rho_0$ its reference density profile. The deformed stellar coordinates $X^i$ and density $\rho$ obey
    \begin{equation}\label{eq:affine_volume_map}
      X^i=\frac{a_i}{R_0}x_0^i,\qquad
      \rho\,d^3X=\rho_0\,d^3x_0,
    \end{equation}
    with the internal velocity supplied by the circulation variable
    \cite{CarterLuminet1985,Lai:1994hf}. This normalization gives
    $I_{ij}=\kappa_nm a_i^2\delta_{ij}/5$ and the energies of
    Eq.~\eqref{eq:affine_internal_energies}.
    Differentiating the internal and self-gravitating energies yields
    \begin{align}
      \frac{\partial U}{\partial a_i}&=-\frac{U}{na_i},\notag\\
      \frac{\partial W_{\rm self}}{\partial a_i}
        &=\frac{3m^2a_i}{2(5-n)}
          \int_0^\infty\frac{du}{(a_i^2+u)
                             \sqrt{\prod_j(a_j^2+u)}}.
      \label{eq:affine_variations}
    \end{align}
    These coefficients reproduce the force terms in
    Eqs.~\eqref{eq of motion 1}--\eqref{eq of motion 3}; at the
    spherical limit the equations reduce to exact hydrostatic balance.
    The conserved internal circulation is
    \begin{equation}\label{eq:circulation_vorticity}
       \mathcal C=\frac{\kappa_nm}{5}a_1a_3
                 (\nabla\times\boldsymbol v)_2,
    \end{equation}
    giving Eq.~\eqref{eq:zero_circulation} on the irrotational
    branch. The figure rotation is measured relative to the
    parallel-transported frame rather than coordinate time.

    \section{Shape coordinates and static fold reduction}
    \label{app:fold}
    Define $\boldsymbol b=(\bar a_1,\bar a_2,\bar a_3)$ and
    $\boldsymbol b=\boldsymbol h(\boldsymbol y)
                   =(y_1,y_2,y_1y_3)$.
    Let $\boldsymbol F_b(\boldsymbol b,r)$ be the static force balance residual expressed in terms of the three semi-axes, and let $J_b=\partial\boldsymbol F_b/\partial\boldsymbol b$ be its Jacobian at fixed orbital radius. At fixed $r$, the residual in shape coordinates is
    $\boldsymbol F(\boldsymbol y,r)
      =\boldsymbol F_b(\boldsymbol h(\boldsymbol y),r)$, so that
    \begin{equation}
        J_y=J_b\frac{\partial\boldsymbol b}{\partial\boldsymbol y},
        \qquad
        \frac{\partial\boldsymbol b}{\partial\boldsymbol y}
         =\begin{pmatrix}1&0&0\\0&1&0\\x&0&\bar a_1\end{pmatrix}.
        \label{eq:shape_chain_jacobian}
    \end{equation}
    The change of coordinates has positive Jacobian determinant
    $\bar a_1$ and thus preserves fold rank. At an ordinary fold,
    $J_y$ has rank two. Let $\boldsymbol v$ and $\boldsymbol u$
    be normalized right and left null vectors, respectively, so
    that $J_y\boldsymbol v=0$ and $\boldsymbol u^TJ_y=0$.
    The transversality and quadratic nondegeneracy coefficients are
    \begin{equation}
      \gamma_{\rm fold}=\boldsymbol u^T\partial_r\boldsymbol F\ne0,
      \qquad
      \beta_{\rm fold}=\boldsymbol u^TD_y^2\boldsymbol F
                    [\boldsymbol v,\boldsymbol v]\ne0.
      \label{eq:fold_nondegeneracy}
    \end{equation}
    Here $\partial_r\boldsymbol F$ includes radial variation of
    the external tidal field and the frame rotation rate, while the
    Hessian is evaluated at fixed radius.
    The second derivative obeys the full chain rule,
    \begin{equation}
        D_y^2\boldsymbol F[\boldsymbol v,\boldsymbol v]
        =D_b^2\boldsymbol F_b[\boldsymbol b_{,y}\boldsymbol v,\boldsymbol b_{,y}\boldsymbol v]
             +J_b(0,0,2v_1v_3)^T.
        \label{eq:shape_hessian_chain}
    \end{equation}
    At a fold, $\boldsymbol u^TJ_b=0$ follows from invertibility of
    $\boldsymbol b_{,y}$, so the extra term vanishes only after left-null projection.
    The nonzero or zero character of the projected quadratic
    coefficient is preserved; its numerical value depends on the
    normalization of the null coordinate.

    Writing $\eta$ for the local displacement amplitude along the right null direction $\boldsymbol v$, the Lyapunov--Schmidt reduction gives
    \begin{equation}\label{eq:fold_reduced_appendix}
      0=\gamma_{\rm fold}\Delta r
       +\tfrac12\beta_{\rm fold}\eta^2
       +O(\eta^3,\eta\Delta r,\Delta r^2),
    \end{equation}
    with the nonzero coefficients defined in
    Eq.~\eqref{eq:fold_nondegeneracy} \cite{Kuznetsov2004}.
    On the side where $-2\gamma_{\rm fold}\Delta r/\beta_{\rm fold}>0$,
    the leading local equation has two solutions,
    $\eta\simeq\pm\sqrt{-2\gamma_{\rm fold}\Delta r/\beta_{\rm fold}}$.
    These represent the two local arms of an ordinary static fold;
    the reduction alone does not identify which arm belongs to the
    nearly spherical sequence or establish a separately continued
    numerical branch.

    Parameter continuation in $r$ alone becomes singular at a fold.
    A separate local arclength parameter $s$ used for numerical continuation instead satisfies
    \cite{AllgowerGeorg1990,Seydel2010,DicksonEtAl2007}
    \begin{equation}
        J_y\frac{d\boldsymbol y}{ds}
           +\boldsymbol F_{,r}\frac{dr}{ds}=0.
        \label{eq:continuation_tangent}
    \end{equation}
    The bordered system
    \begin{equation}\label{eq:fold_bordered_matrix}
      \begin{pmatrix}
        J_y&\boldsymbol F_{,r}\\\boldsymbol v^T&0
      \end{pmatrix}
    \end{equation}
    is nonsingular at an ordinary fold. We use Jacobian rank,
    transversality, quadratic nondegeneracy, and independent
    weak-tide branch continuation as separate numerical diagnostics
    \cite{Seydel2010}.

    \section{Polytropic WD normalization and units}
    \label{app:WD_units}
    The $n=3/2$ WD model adopts a nonrelativistic degenerate-electron
    EOS \cite{Chandrasekhar1939}. Relativistic degeneracy,
    Coulomb effects, and corrections that depend on composition require
    different stellar models \cite{Nauenberg1972,HamadaSalpeter1961,Salpeter1961,RotondoEtAl2011}.
    With the Lane--Emden variables of Eq.~\eqref{eq:Lane_Emden},
    the normalized second moment is
    \begin{equation}\label{eq:LE_second_moment_normalization}
       \frac{I_{ii}}{mR_0^2}
         =\frac{1}{3\xi_1^2\nu_n}
           \int_0^{\xi_1}\Theta^n\xi_{\rm LE}^4d\xi_{\rm LE}
         =\frac{\kappa_n}{5}\quad\text{(no sum)}.
    \end{equation}
    The physical dynamical time is
    $t_*^{\rm SI}=\sqrt{R_0^3/(Gm_{\rm SI})}$, with
    geometric time $t_*=ct_*^{\rm SI}$. Mass density and pressure
    transform as $\rho_{\rm geo}=G\rho/c^2$ and
    $P_{\rm geo}=GP/c^4$. To apply the black hole backgrounds to the WD
    equations, let $M_{\rm ADM,code}$ be the ADM mass in the black hole
    solver length units and define $L_{\rm code}=M_{\rm ADM}/M_{\rm ADM,code}$,
    with physical masses expressed as geometrized lengths. The conversions are
    \begin{align}
      \bar R_0&=\frac{R_0}{m^{1/3}M_{\rm ADM}^{2/3}},\notag\\
      \bar c_{ij}&=\frac{R_0^3}{mL_{\rm code}^2}c_{ij,\rm code},\notag\\
      \bar\omega_{\rm fr}&=\sqrt{\frac{R_0^3}{mL_{\rm code}^2}}\omega_{\rm fr,code},\notag\\
      M_{\rm ADM}\Omega_{\rm orb}&=M_{\rm ADM,code}\Omega_{\rm orb,code}.
      \label{eq:BH_WD_unit_conversion}
    \end{align}
    Here $\bar\omega_{\rm fr}=t_*\omega_{\rm fr}$,
    consistent with the aligned stellar condition. Curvature has
    dimensions of inverse length squared and the frame rate of inverse length.
    At fixed composition, the model has
    $R_0\propto m^{-1/3}$. Here $p_F=\hbar[3\pi^2\rho_c/(\mu_e m_u)]^{1/3}$ is the electron Fermi momentum at the stellar center. The value $p_F/(m_ec)\simeq0.694$ illustrates the limitation of a strictly nonrelativistic EOS.

\end{appendices}

\bibliographystyle{unsrt}
    \bibliography{0_MyRef}

\end{document}